\documentclass[%
 reprint,
 amsmath,amssymb,
 aps,
]{revtex4-2}

\usepackage{graphicx}
\usepackage{dcolumn}
\usepackage{bm}
\usepackage{xcolor}
\usepackage{float}
\usepackage{hyperref}

\begin{document}

\preprint{APS/123-QED}

\title{Propagation of circular Airy derivative beams in complex media}
\author{Anita Kumari}
\author{Vasu Dev}
\author{Vishwa Pal}%
 \email{vishwa.pal@iitrpr.ac.in}
\affiliation{Department of Physics, Indian Institute of Technology Ropar, Rupnagar, Punjab 140001, India}%

\date{\today}

\begin{abstract}
Controlling light propagation through complex media plays a significant role in a wide range of applications ranging from astronomical observations to microscopy. Although, several advances have been made based on adaptive optics, optical phase conjugation and wavefront shaping, but many of these involve challenges. Recently, controlling light propagation in complex media by simply structuring light has shown promising capabilities. We present experimental and numerical investigations of abruptly autofocusing property of circular Airy derivative beams (CADBs) in complex media. We find that up to a relatively high turbulence strength, CADB possesses relatively good abrupt autofocusing, however, efficiency and autofocusing position (longitudinal and transverse) vary with the strength of turbulence. Further, the spatial distortions in CADB caused by turbulence are quantified by an overlap integral, which shows that CADB possesses reasonably good resilience against the turbulence. The diffraction efficiency of CADB changes by a factor of $\sim 1.7$ with increasing strength of turbulence from zero to high, indicating good confinement of intensity at autofocusing. The focused beam spot size grows gradually with increasing the strength of turbulence, specifically, it grows by a factor of $\sim 2$ for a strong turbulence, indicating reasonably good focusing abilities. The results of CADB are compared with a Gaussian beam, and find that CADB possesses superior focusing abilities in turbulent media. We have carried out a detailed analysis of these observations based on Zernike polynomials, which reveals that different kinds of aberrations present in turbulent media leads to distortions in the spatial structure as well as other properties of CADBs. Our results can be used for various applications, such as in biomedical treatment, seismology, optical tweezers and material processing.
\end{abstract}

\maketitle

\section{Introduction}
\label{intro}
The propagation of light through complex media (e.g., turbulence or biological tissue) causes various deleterious effects, which pose strong limitations in various applications, such as in optical communications, biomedical imaging, seismology, optical tweezers, and material processing \cite{ren2021,papazoglou2011,yan2018}. In particular, when light travels through a random media (turbulence), it experiences significant impairment due to the presence of different types of aberrations which produce distortions in both phase and amplitude of the light beam. This phenomenon gives rise to undesirable effects such as increased beam dispersion, fluctuations in intensity, coherence loss, beam deviation, and scintillation \cite{lagendijk1996,schwartz2008,lazer2023}. Therefore, controlling light propagation in complex media has been the subject of much attention to overcome these adverse effects.

To tackle these challenges, various efforts have been made based on adaptive optics \cite{Tyson2010}, holography and phase conjugation \cite{Leith1966,HE2002,Wang2015}, and wavefront shaping techniques \cite{Vellekoop2007,Cao2022,Pai2021,Katz2011,Nixon2013}. Among these, the wavefront shaping techniques have gained wide attention, and several exciting investigations have been performed over the years. One of pioneering works that triggered the field includes the focusing of light through a strong scattering media by shaping the incident wavefront with a spatial light modulator (SLM) \cite{Vellekoop2007}. The wavefront shaping techniques have also been exploited for imaging and transmission \cite{Yaqoob2008,Popoff2010}, controlling the scattered light in both space and time \cite{Katz2011,Aulbach2011}, and controlling its spectral and polarization properties \cite{Guan2012,Small2012}. Since wavefront shaping techniques rely on SLM with electronic feedback loop \cite{Vellekoop2007}, so possesses inherently slow response times \cite{Conkey2012}. A rapid all-optical approach based on self-organization of optical field inside a multimode laser cavity has also been demonstrated for focusing light scattered by rapidly varying inhomogeneous media \cite{Nixon2013}. However, it involves a complex design of experimental setup.

Further, it has been shown that increasing the complexity of the beam can lead to an improvement in the transmission performance of beams in turbulent media \cite{gianani2020}. Researchers have consequently proposed various beam types and thorough investigations have been conducted to examine the impact of turbulence on their different properties with the goal of developing systems that can compensate their undesired effects, such as the scintillation behavior of Laguerre Gaussian beams in strong turbulence \cite{eyyubouglu2011}, propagation of higher-order Bessel-Gaussian beams in turbulence \cite{eyyubouglu2007}, long-distance Bessel beam propagation through Kolmogorov turbulence \cite{birch2015}, coherence of Bessel beam in a turbulent atmosphere \cite{lukin2012}, propagation of combining Airy beam in turbulence \cite{wen2015}, beam wander of partially coherent Airy beam in oceanic turbulence \cite{jin2018}, influences of atmospheric turbulence effects on the orbital angular momentum spectra of vortex beams \cite{fu2016}, synthesized vortex beams in the turbulent atmosphere \cite{aksenov2020}, partially coherent elegant Hermite-Gaussian beam in turbulent atmosphere \cite{wang2011}, autofocusing and self-healing properties of Aberration laser beams in a turbulent media \cite{dev2021}, and propagation of vortex symmetric Airy beam in turbulent link \cite{zhang2023}. Recently, the invariance and distortion of vectorial light across a real-world free space link has also been investigated \cite{Peters2023}. The field of structuring light has also been extended to the quantum domain for controlling the propagation of photons in complex media \cite{Lib2022}. A few recent investigations includes the generation of quantum Airy photons \cite{Maruca2018} and  spatially entangled Airy photons \cite{Lib2020}.

In free space optical communication, symmetric Airy beams emerge as suitable optical carriers for enhancing robustness against atmospheric turbulence. This is attributed to their unique abrupt autofocusing feature, leading to an increased signal-to-noise ratio, a crucial factor in improving optical signal strength \cite{zhang2023,vaveliuk2014}. The influence of various aberrations on the intensity distribution of vortex circular Airy beams (CABs) has been previously examined under different atmospheric turbulent conditions in which they have shown that spherical aberration has the most significant impact. In contrast, z-tilt has the least effect on the intensity distribution of vortex CABs \cite{lazer2023}. Recently, circular Airy derivative beams (CADBs) have been shown to exhibit abrupt autofocusing features and good self-healing against various obstructions in free space \cite{kumari2024,zang2022}. It has been shown that CADBs possess stronger abrupt autofocusing than normal CABs \cite{zang2022}. The k-value of CADBs at its focal plane becomes higher by several factors than CABs under the same conditions, which makes them more suitable for applications \cite{ren2021,porfirev2021}. So far CADBs have been investigated only in free space, so a natural question arises how these beams behave when propagated in complex media, which is highly relevant for their potential use in a wide range of applications. How better is their performance as compared to conventional Gaussian beams?
 
Here, we have investigated numerically and experimentally the abruptly autofocusing of CADBs in turbulence with varying strength, and performed a depth analysis as well as their performance comparison with conventional Gaussian beams. Further, to understand the deleterious effects caused by turbulence, we have performed a detailed analysis based on Zernike polynomials.

The paper is organized as follows. In Sec.\,\ref{theo_desc}, we present theoretical descriptions of CADBs and method for generating turbulence in the lab. Section\,\ref{exp} presents the experimental setup for generation and propagation of CADBs through turbulent media of different strengths. In Sec.\,\ref{results}, numerical and experimental results are presented for CADBs in turbulent media with weak, moderate and strong turbulence conditions. The quantified results as well as the performance comparison with Gaussian beam are also presented. In Sec.\,\ref{Explan}, we present the physical explanations of various effects caused by turbulent media, based on calculating the Zernike polynomials.  Finally, in Sec.\,\ref{concl}, we provide the concluding remarks.

\section{Theoretical description}
\label{theo_desc}
The electrical field of CADB propagating along $z-$axis at initial plane ($z=0$) can be given as \cite{kumari2024,zang2022} 
\begin{equation} 
 E(r,\theta)=\ Ai^{(n)}\Big(\frac{r_0 - r}{w_0}\Big) \exp\Big(a\frac{r_0 - r}{w_0}\Big), ~~~~~~r\leq R\label{eq1}
\end{equation}
where $r=\sqrt{(x^2+y^2)}$ and $\theta=\tan^{-1}\left(y/x\right)$ represent the radial and azimuthal coordinates, respectively. $a$ is an exponential decay factor, $\ w_0$ is a scaling factor, $\ r_0$ is the parameter related to the radius of the primary ring of CADB at $z=0$ and $R$ denotes the mean screen radius. $\ Ai^{(n)}(^.)$ is $n^{\mathrm{th}}$-order derivative of the Airy function with respect to $r$. $n$ is the order of derivative and non-negative integer. 
\begin{figure}[htbp]
\centering
\includegraphics[height = 4.0cm, keepaspectratio = true]{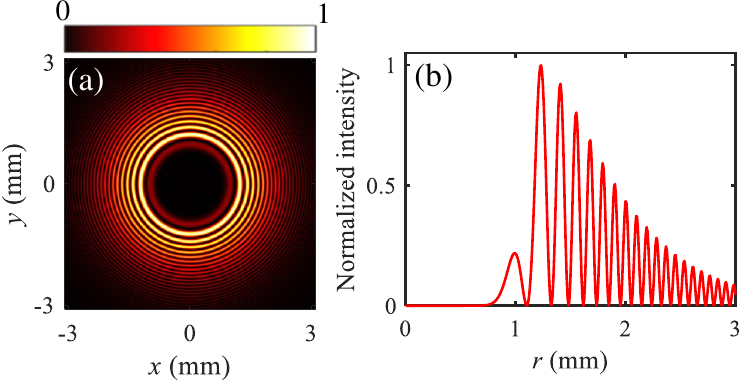}
\caption{(a) Normalized intensity distribution of CADB at initial plane $z=0$. (b) Intensity cross-section taken along the radial direction in (a).}
\label{fig1}
\end{figure} 

The intensity distribution of CADB at an initial plane $z=0$ is shown in Fig.\,\ref{fig1}(a), indicating multiple concentric ring-like structure. Figure.\,\ref{fig1}(b) shows the intensity cross-section along the radial direction, indicating an exponential decay of intensity along the radial direction. The autofocusing ability of CADBs is quantified by an intensity contrast referred to as the k-value. It is defined as $I_{\mathrm{max}}(z>0)/I_{\mathrm{max}}(z=0)$, where $I_{\mathrm{max}}(z>0)$ and $I_{\mathrm{max}}(z=0)$ denote the maximum intensities observed at $z>0$ and $z=0$ planes, respectively \cite{kumari2024}. The CADB possesses autofocusing due to lateral self-acceleration, where measured k-value consistently reaches its maximum at the autofocusing point ($z_{af}$)\cite{kumari2024}. The abruptness in autofocusing can be quantified by measuring the full-width at half-maximum (FWHM) of the k-value variation with propagation distance $z$, which determines how fast k-value changes around $z_{af}$. It has been shown that the abrupt autofocusing increases with the order $n$ \cite{zang2022}, and we have considered CADBs with $n=1$. The abrupt autofocusing in CADBs also depends on various other beam parameters, such as $r_o$, $a$, and $w_o$ \cite{zang2022}. In this work, our aim is to investigate the propagation of CADBs in the turbulent media with varying strength, and to explore their abrupt autofocuisng properties. 

The theory of turbulent air is based on temperature differences across different points in the atmosphere. The impact of these small temperature fluctuations leads to variations in the refractive index of the air. These fluctuations originate initially in phase, but after propagation, they translate to both phase and amplitude of optical fields, resulting in distortions in the field. The Kolmogorov model, which we have considered in our investigations, can able to establish a correlation between temperature fluctuations and fluctuations in the refractive index of the atmosphere \cite{cox2020}. In this paper, the $N \times N$ size phase screen with the Kolmorgov spectrum model $\Phi_{n}(k)$ is used to simulate turbulence, which is given by
\begin{equation} 
\phi(x,y)=\ F^{-1}(R_{NN}\sqrt{\Phi_{n}(k)}), ~~~~~~\label{eq2}
\end{equation}
where $F^{-1}$ denotes the inverse Fourier transform, $R_{NN}$ represents a complex random matrix having a normal distribution with zero mean and unit variance. The Kolmorgov spectrum model $\Phi_{n}(k)$ is given by \cite{cox2020} 
\begin{equation} 
\Phi_{n}(k)=0.033~C{_n}^2 ~k^{-11/3}, ~~~~~~~~1/L_o<<k<<1/l_o  \label{eq3}
\end{equation}
where $k = 2\pi f$ is the spatial angular frequency and $ C{_n}^2$ is the refractive index structure parameter. $L_o$ and $l_o$ are the outer and inner scale of turbulence and for the simplification of the model, these are assumed to be $\infty$ and $0$, respectively. Here, we considered the thin-phase screen approximation in which using only one phase screen which is good enough because the considered length, $L = 3.32$ m is not too large \cite{burger2008}. Within this approximation, the turbulence strength is completely characterized by a dimensionless parameter, denoted as $D/r_{f}$. Here, $D$ is the diameter of the beam, and $r_{f}$, known as Fried's parameter, is the atmospheric coherence length given as \cite{cox2020} 
\begin{equation} 
r_{f}=\ 1.68~(C{_n}^2Lk_o^{2})^{-3/5}, ~~~~~~\label{eq4}
\end{equation}
where $k_o$ is the vacuum optical wavenumber.
The turbulence strength can be quantified by a less descriptive parameter, the Strehl ratio ($\mathrm{S_R}$). It is defined as the ratio of the average on-axis beam intensity with turbulence $I_{T}$ to that without turbulence $I_{o}$ as \cite{cox2020}
\begin{equation} 
\mathrm{S_R}= \frac{I_{T}}{I_{o}}\approx \frac{1}{1+(D/r_{f})^{5/3}}, ~~~~~~\label{eq5}
\end{equation}
\begin{figure}[ht!]
\centering
\includegraphics[height = 6cm, keepaspectratio = true]{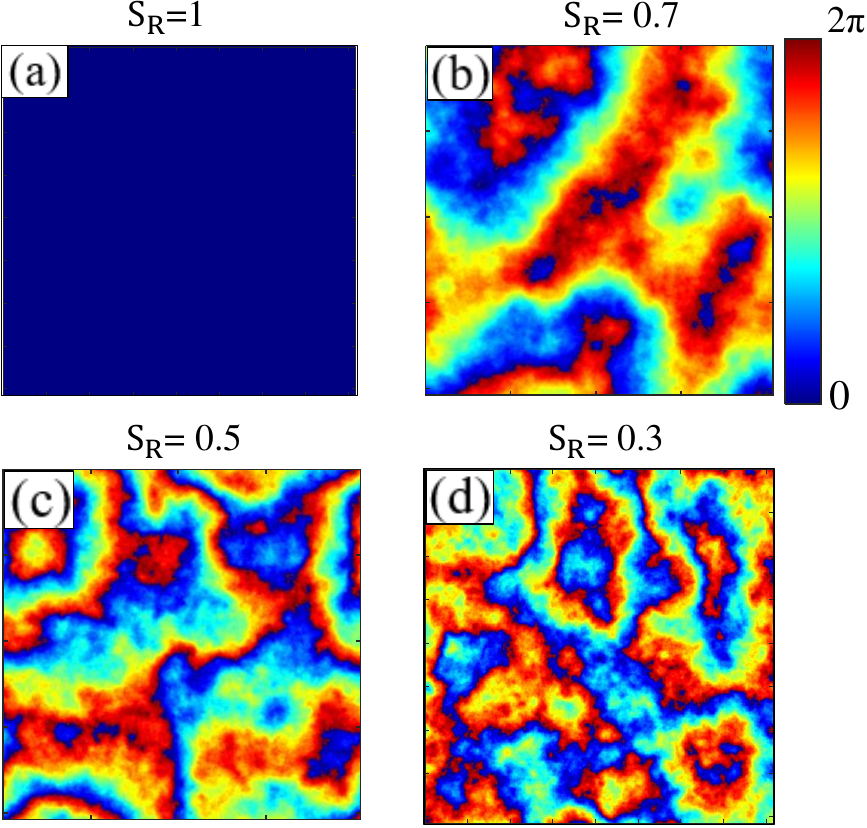}
\caption{The phase screen with (a) $\mathrm{S_R}=1$ (no turbulence), (b) $\mathrm{S_R}=0.7$ (low strength turbulence), (c) $\mathrm{S_R}=0.5$ (medium strength turbulence), and (d) $\mathrm{S_R}=0.3$ (high strength turbulence).}
\label{fig2}
\end{figure}

The value of $\mathrm{S_R}$ lies between $0$ to $1$. The maximum value of $\mathrm{S_R}=1$ denotes that there is no turbulence, i.e., $I_{T}=I_{o}$, and the minimum value of $\mathrm{S_R}=0$ indicates that the medium is highly turbulent, i.e., $I_{T} << I_{o}$. For our investigations, we have considered four different turbulence phase screens with varying strength, such as $\mathrm{S_R}=1$ (no turbulence), $\mathrm{S_R}=0.7$ (low strength turbulence), $\mathrm{S_R}=0.5$ (medium strength turbulence), and $\mathrm{S_R}=0.3$ (high strength turbulence), as shown in Fig.\,\ref{fig2}. As evident, the phase screen with $\mathrm{S_R}=1$ consists of a uniform phase distribution (Fig.\,\ref{fig2}(a)), and with decreasing the values of $\mathrm{S_R}$ the randomness in the phase distribution increases (Figs.\,\ref{fig2}(b)-\ref{fig2}(d)). 

We have numerically simulated the propagation of CADBs in a turbulent media by using an extended Huygens-Fresnel integral as \cite{dev2021,andrews2005}
\begin{eqnarray}
E(r',\theta',z)&=& \frac{k_o}{2\pi iz}\int_{0}^{\infty}\int_{0}^{2\pi} E(r,\theta,z) \nonumber\\
&&\times\exp\left(\frac{ik_o}{2z}[r^2+{r'}^{2} -2rr'\cos(\theta - \theta')]\right)\nonumber \\
&&\times \exp[\psi(r',r,z)] ~ rdr d\theta, \label{eq6}
\end{eqnarray}
where $k_o=2\pi/\lambda$ with $\lambda$ is the optical wavelength, ($r,\theta$) and ($r',\theta'$) are the coordinates of initial ($z=0$) and observation planes ($z>0$), respectively, $\psi(r',r,z)$ represents a random complex phase function. For our investigations, throughout the paper we have chosen the CADB parameters as $r_0=1$ mm, $w_0=0.1$ mm, $\lambda=1064$ nm, $n=1$, and $a=0.1$ \cite{zang2022}. 
\section{Experimental arrangement} \label{exp}
\begin{figure*}[ht!]
\centering
\includegraphics[height = 10.12cm, keepaspectratio = true]{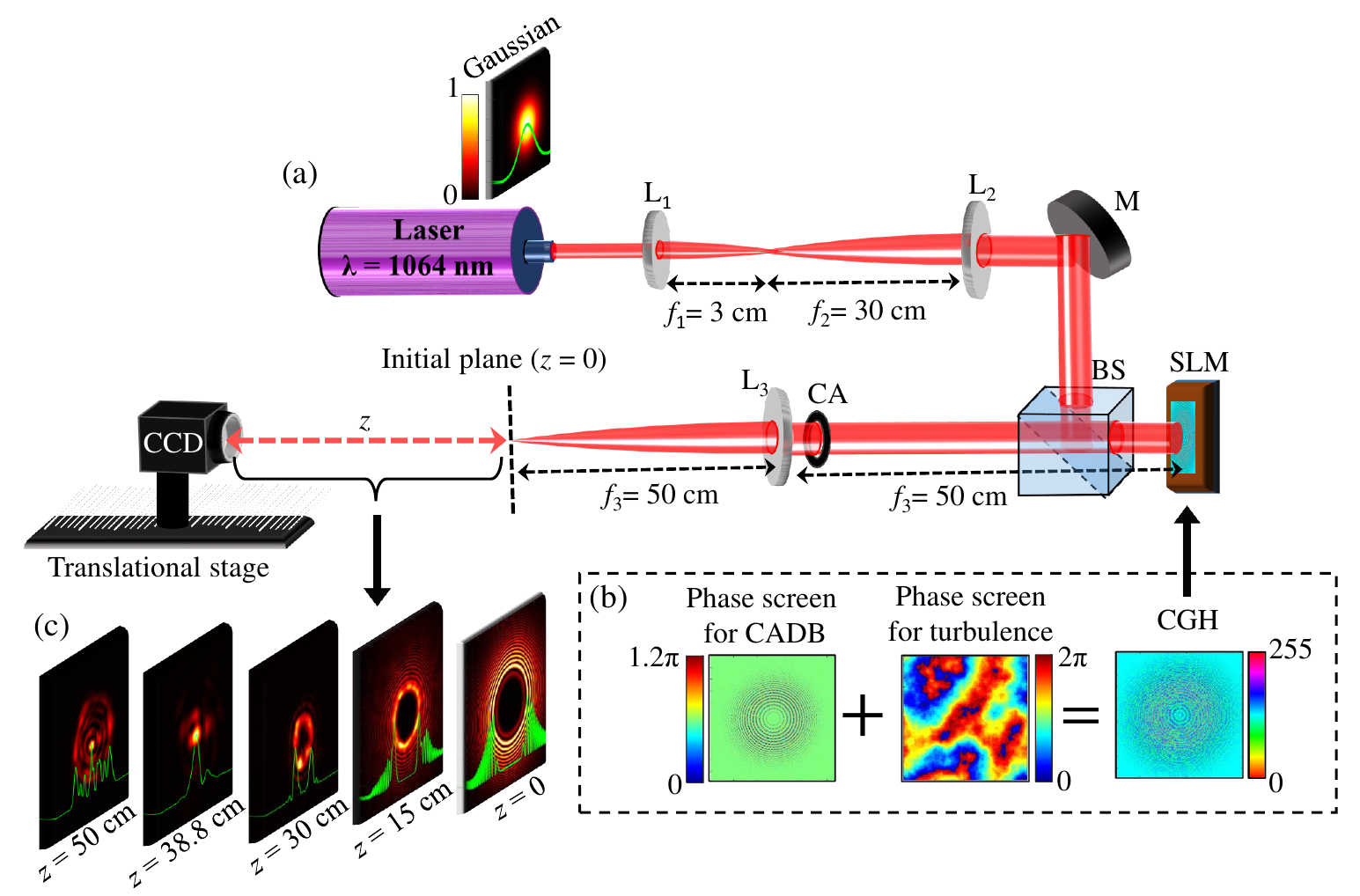}
\caption{(a) Experimental arrangement for generation and propagation of CADBs in a turbulent media. (b) The computer generated hologram (CGH) formed by adding the phase distributions corresponding to CADB and turbulence. The phase of CGH is converted to 256 levels for applying onto SLM screen. (c) Intensity distribution of CADB at different propagation distances, indicating the spatial distortions as well as autofocusing. L$_1$, L$_2$ and L$_3$: Plano-convex lenses of focal length $f_1=3$ cm,$f_2=30$ cm and $f_3=50$ cm, respectively; M: mirror; BS: 50:50 beam splitter; SLM: spatial light modulator; CA: circular pinhole; CCD: charge-coupled device.}
\label{fig3}
\end{figure*} 
The experimental arrangement for the generation and propagation of CADBs in turbulent media is shown in Fig.\,\ref{fig3}(a). A linearly polarized Gaussian laser beam at a wavelength of $\lambda$ = 1064 nm is magnified with two planoconvex lenses $L_1$~($f_1=3$ cm) and $L_2$~($f_2=30$ cm), and then incidents perpendicularly on a spatial light modulator (SLM) with the help of a 50:50 beam splitter (BS). The SLM consists of screen resolution $1920\times 1080$ and each pixel of size $8 ~\mu$m. The amplitude and phase of an incident laser beam are modulated by imposing a suitable phase pattern onto the SLM. The modulated light consists of several orders, and a suitable first order is selected by an aperture (CA). A planoconvex lens $L_3$~($f_3=50$ cm) performs the Fourier transform, and desired CADB is obtained at the back focal plane of $L_3$ (marked as an initial plane $z=0$ in Fig.\,\ref{fig3}(a)). The propagation of CADB is recorded with a CCD (charged-coupled device) mounted on a translation stage.

To investigate the propagation of CADBs in a turbulent media, we have simulated turbulence in the laboratory using SLM \cite{burger2008,bold1998,rosalesshape2017}. Particularly, a turbulent phase screen is generated using the method described above, which behaves like a turbulent medium of desired length and turbulence strength, and imposed onto the SLM. 
This method provides increased flexibility compared to real dynamic turbulence, as the turbulence model can be altered on demand as well as the parameters such as strength and length of turbulence. 

The complex Fourier-transformed electric field of CADBs (Eq.\,(\ref{eq1})) can be expressed as 

\begin{equation}
E(\rho,\phi')=F[E(r,\theta,0)]=A(\rho,\phi')\exp[i\zeta(\rho,\phi')],\label{eq7}
\end{equation}
where $F$ denotes the Fourier transform operation, and amplitude $A(\rho,\phi')$ is normalized to a maximum value of $1$. Now we encode the complex field $E(\rho,\phi')$ using a phase transmittance function (phase-only computer-generated hologram) to incorporate the amplitude and phase variations as
\begin{equation}
 T(x,y)=\exp[i(\Psi(A,\zeta)],\label{eq8}
\end{equation}
where $\Psi(A,\zeta)$ takes into account the amplitude and phase variations corresponding to a complex field $E(\rho,\phi')$ (as shown in Fig.\,\ref{fig3}(b)), and we calculate it using a Fourier series based method \cite{kumari2024,arrizon2007}. Further, to account the effect of turbulence on CADBs, a random phase distribution corresponding to a desired turbulence is added to the phase distribution corresponding to the complex field $E(\rho,\phi')$ as {\cite{rosalesshape2017}}
\begin{equation}
 T(x,y)=\exp[i(\Psi(A,\zeta + \psi(r',r,z))],\label{eq9}
\end{equation}
The transmittance $T(x,y)$ contains several higher orders along with a desired first order \cite{kumari2024,arrizon2007}. To separate the desired order, a blazed grating is added to the phase pattern. The final modified and required phase pattern can be written as $\Psi(A,\zeta+2\pi f_x x+2\pi f_y y+\psi(r',r,z))$, where $f_x$ and $f_y$ are the spatial frequencies along $x-$ and $y-$directions, respectively. Both the spatial frequencies $f_x$ and $f_y$ are optimized and chosen to be the same as $1300$ mm$^{-1}$, in the experiment. The final phase pattern is converted into 256 levels for imposing it on the SLM screen which produces CADBs with distortions corresponding to the desired strength of a turbulent media, as shown in Fig.\,\ref{fig3}(c).
\section{Results and discussion}\label{results}
To show the resilience and abrupt autofocusing abilities of CADBs, we have numerically and experimentally investigated the propagation of these beams through the turbulent media of varying strength, namely, low, medium and high strength. To quantify the effect of turbulence on the properties of CADBs, we have compared the results with the CADB in free space.
\begin{figure*}[htbp]
\centering
\includegraphics[height = 9.40 cm, keepaspectratio = true]{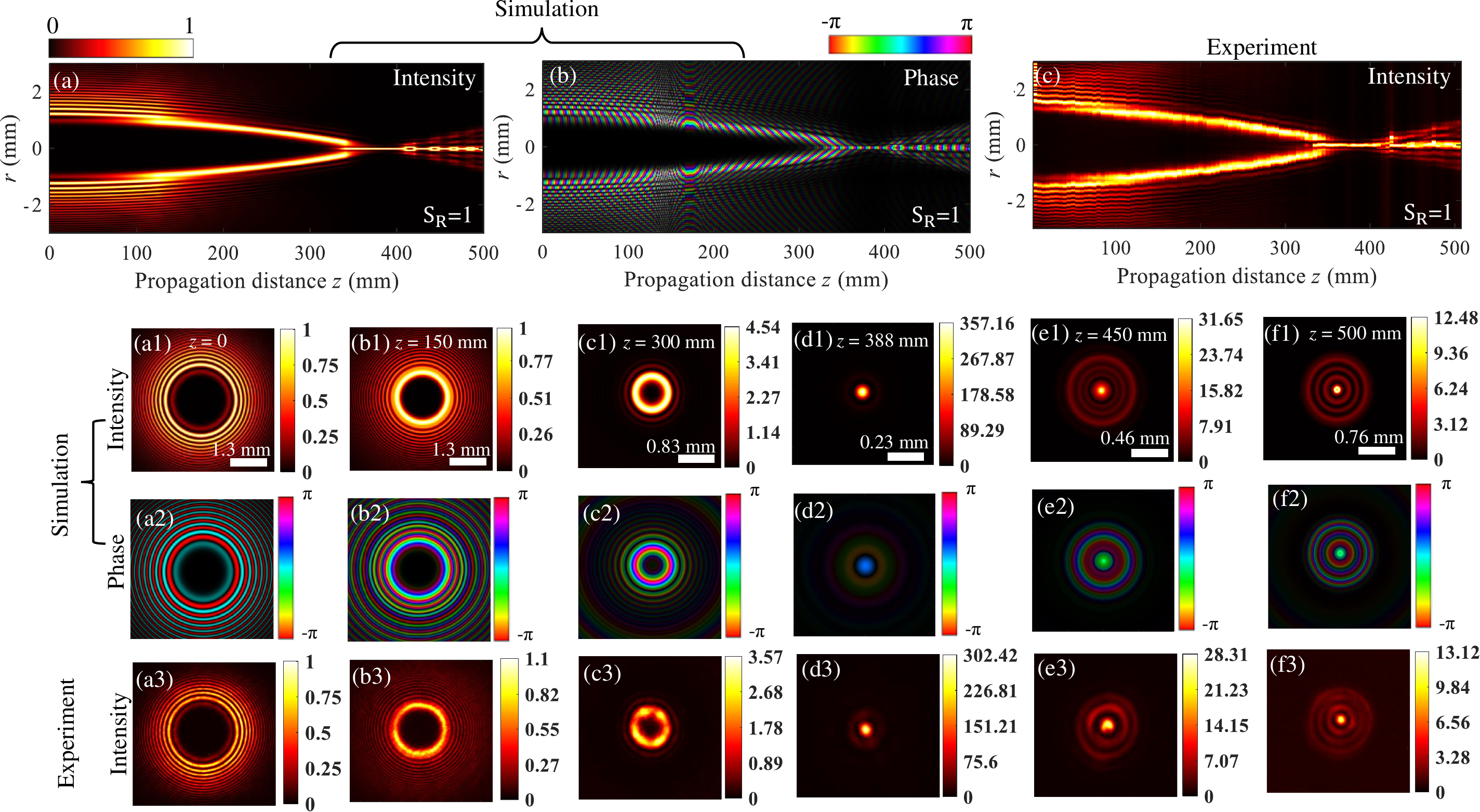}
\caption{In a free space ($\mathrm{S_R}=1$), the intensity (a, c) and phase (b) distributions of CADB as a function of propagation distance $z$.  The snapshot of intensity and phase distributions of CADB at propagation distance (a1, a2, a3) $z=0$, (b1, b2, b3) $z=150$ mm, (c1, c2, c3) $z=300$ mm, (d1, d2, d3) $z=388$ mm, (e1, e2, e3) $z=450$ mm, and (f1, f2, f3) $z=500$ mm. The other parameter values are taken as $r_0=1$ mm, $w_0=0.1$ mm, $a=0.1$ and $\lambda=1064$ nm.}
\label{fig4}
\end{figure*} 

The simulated and experimental results of propagation of CADB in free space ($\mathrm{S_R}=1$) are shown in Fig.\,\ref{fig4}. Figures\,\ref{fig4}(a) and \ref{fig4}(b) show the simulated results of evolution of cross-sectional intensity and phase of CADB as a function of propagation distance. Figure\,\ref{fig4}(c) shows the experimental results of evolution of cross-section intensity of CADB as a function of propagation distance. Figures\,\ref{fig4}(a1)-\ref{fig4}(f1), Figs.\,\ref{fig4}(a2)-\ref{fig4}(f2) and Figs.\,\ref{fig4}(a3)-\ref{fig4}(f3) show the simulated intensity, simulated phase and experimental intensity distributions of CADB at different propagation distances, respectively.  Note, in Figs.\,\ref{fig4}(a1)-\ref{fig4}(f1) and Figs.\,\ref{fig4}(a3)-\ref{fig4}(f3), the highest value on the colorbar denotes the k-value at the respective propagation distance. 

As evident, initially at $z=0$ CADB consists of multiple concentric bright rings with a relative phase of $\pi$ between the neighbouring bright rings, and as the CADB propagates due to lateral self-acceleration the adjacent rings begin to interfere from the outermost to the inner, and because of this intensity redistribution takes place. As a result, the number of multiple rings gradually decreases, and eventually, at a distance of $z = 388$ mm a major part of intensity becomes tightly focused in the form a high-intensity peak (Figs.\,\ref{fig4}(d1) and \ref{fig4}(d3)), referred to as the autofocusing distance ($z_{af}$). The autofocusing behaviour in CADBs has also been explained by an analogy to the Fresnel zone plate that composed of the specially designed transparent concentric rings \cite{zang2022}. After $z_{af}$, the intensity again redistributes in the form of multiple bright rings. As shown, with increasing the propagation distance the k-value gradually increases from 1 to 4.5 until $z\approx300$ mm, and after that it increases abruptly to a maximum value of $\sim 357$ in the simulation (Figs.\,\ref{fig4}(d1)) and $\sim 303$ in the experiment (Figs.\,\ref{fig4}(d3)) at $z_{af}=388$ mm. After $z_{af}$, the k-value again drops abruptly and then continues decreasing gradually. Note, a small discrepancy between the simulated and experimental k-value is anticipated by a slight inaccurate modulation of the tail part of the input Gaussian beam on the SLM, which affects the outer low-intensity rings of generated CADB. The outer-low intensity rings play a significant role in the focusing efficiency of CADB, and thereby affects the k-value \cite{kumari2024}. This variation in the k-value with $z$ occurs due to the redistribution of intensity. A rapid change of k-value around $z_{af}$ evidences an abruptly autofocusing behaviour of CADB.

\begin{figure*}[htbp]
\centering
\includegraphics[height = 9.40cm, keepaspectratio = true]{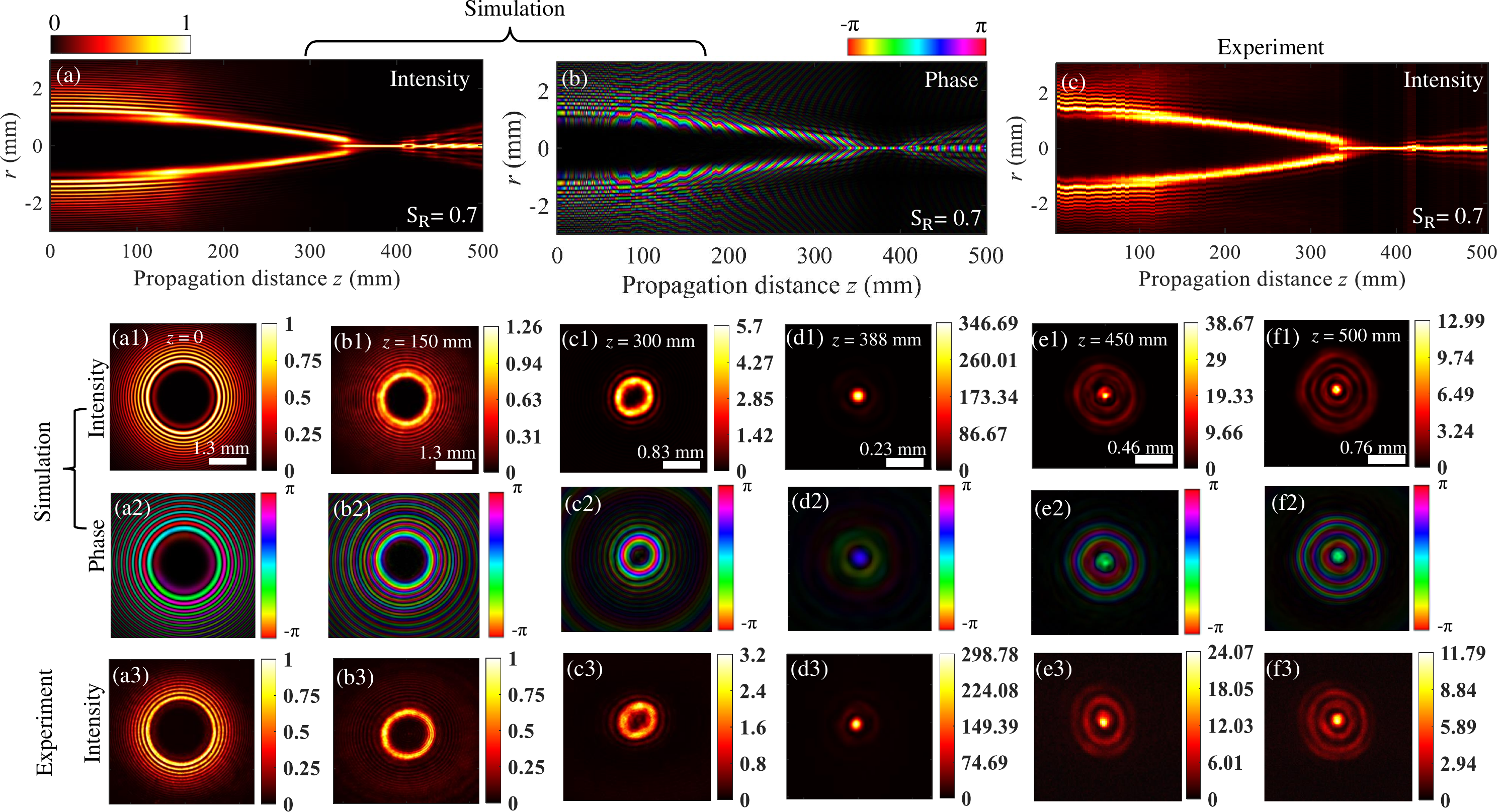}
\caption{In a turbulent media with $\mathrm{S_R}=0.7$, the intensity (a, c) and phase (b) distributions of a CADB as a function of propagation distance $z$. The snapshot of intensity and phase distributions of a CADB at propagation distance (a1, a2, a3) $z=0$, (b1, b2, b3) $z=150$ mm, (c1, c2, c3) $z=300$ mm, (d1, d2, d3) $z=388$ mm, (e1, e2, e3) $z=450$ mm, and (f1, f2, f3) $z=500$ mm. The other parameter values are taken as $r_0=1$ mm, $w_0=0.1$ mm, $a=0.1$ and $\lambda=1064$ nm.}
\label{fig5}
\end{figure*}

In order to analyze the effect of turbulence on the focusing abilities of CADBs, first we have propagated CADB in a low strength turbulent medium with $\mathrm{S_R}=0.7$. The simulated and experimental results are shown in Fig.\,\ref{fig5}. Figures\,\ref{fig5}(a) and \ref{fig5}(b) show the simulated results of evolution of cross-sectional intensity and phase of CADB as a function of propagation distance. Figure\,\ref{fig5}(c) shows the experimental results of evolution of cross-sectional intensity of CADB as a function of propagation distance. Figures\,\ref{fig5}(a1)-\ref{fig5}(f1), Figs.\,\ref{fig5}(a2)-\ref{fig5}(f2) and Figs.\,\ref{fig5}(a3)-\ref{fig5}(f3) show the simulated intensity, simulated phase and experimental intensity distributions of CADB at different propagation distances, respectively. As evident, the low strength turbulence (consists of small randomness in the phase distribution (Fig.\,\ref{fig2}(b)) causes relatively small spatial distortions in the intensity distribution of CADB during the propagation, and the beam still shows good autofocusing behaviour. The autofocusing distance is found to be the same as in the case of free space ($z_{af}=388$ mm). However, the maximum k-value is found to be reduced due to the presence of turbulence, for example, $\sim 347$ in simulation (Fig.\,\ref{fig5}(d1)) and $\sim 299$ in experiment (Fig.\,\ref{fig5}(d3)), indicating a slight decrease in the autofocusing efficiency. Further, with increasing the propagation distance the k-value gradually increases from $1$ to $5.7$ up to a distance $z\approx300$ mm, and after that it abruptly increases to a maximum value of $\sim347$ at $z_{af}=388$ mm (Fig.\,\ref{fig5}(d1)), and after $z_{af}$ it again drops abruptly and then continues decreasing gradually. This variation of k-value with the propagation distance shows the similar trend as observed in the case of free space. This confirms that the abrupt autofocusing behaviour of CADB remains unchanged for low strength turbulent media.

\begin{figure*}[htbp]
\centering
\includegraphics[height = 9.40cm, keepaspectratio = true]{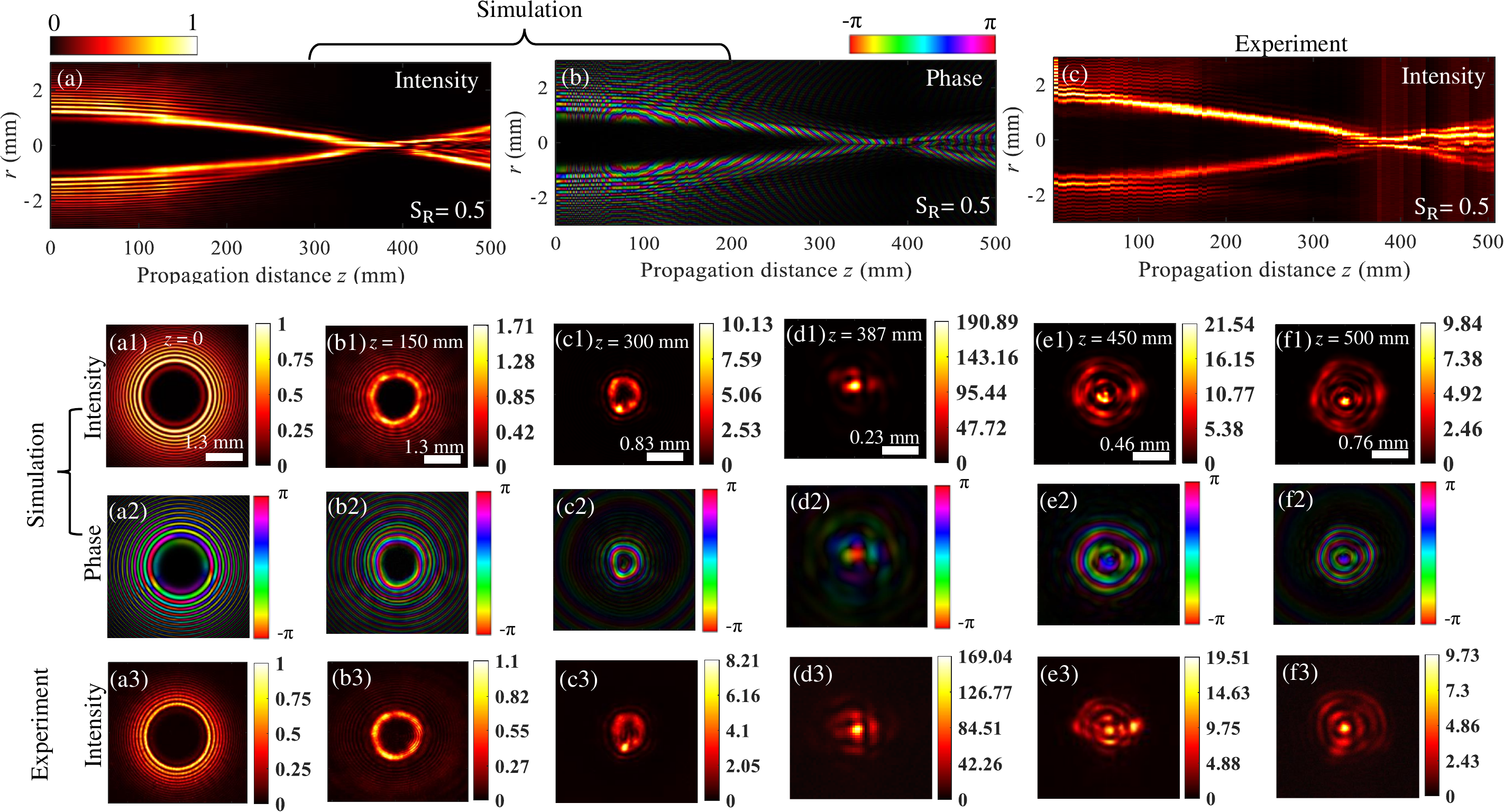}
\caption{In a turbulent media with $\mathrm{S_R}=0.5$, the intensity (a, c) and phase (b) distributions of a CADB as a function of propagation distance $z$. The snapshot of intensity and phase distributions of a CADB at propagation distance (a1, a2, a3) $z=0$, (b1, b2, b3) $z=150$ mm, (c1, c2, c3) $z=300$ mm, (d1, d2, d3) $z=387$ mm, (e1, e2, e3) $z=450$ mm, and (f1, f2, f3) $z=500$ mm. The other parameter values are taken as $r_0=1$ mm, $w_0=0.1$ mm, $a=0.1$ and $\lambda=1064$ nm.}
\label{fig6}
\end{figure*}
We have further increased the strength of turbulence ($\mathrm{S_R}=0.5$), and propagated the CADB through it. The simulated and experimental results are shown in Fig.\,\ref{fig6}. Figures\,\ref{fig6}(a) and \ref{fig6}(b) show the simulated results of evolution of cross-sectional intensity and phase of CADB as a function of propagation distance. Figure\,\ref{fig6}(c) shows the experimental results of evolution of CADB's cross-sectional intensity as a function of propagation distance. Figures\,\ref{fig6}(a1)-\ref{fig6}(f1), Figs\,\ref{fig6}(a2)-\ref{fig6}(f2) and Figs.\,\ref{fig6}(a3)-\ref{fig6}(f3) show the simulated intensity, simulated phase and experimental intensity  distributions of CADB at different propagation distances, respectively. As evident, since the turbulent medium with $\mathrm{S_R}=0.5$ consists of more randomness in the phase distribution (Fig.\,\ref{fig2}(c) and Fig.\,\ref{fig6}(a2)), therefore, upon propagation through such a medium CADB experiences more spatial distortions in the intensity as compared to the previous cases of turbulent media with $\mathrm{S_R}=0.7$ and $\mathrm{S_R}=1$. Due to the randomness in the phase distribution (Figs.\,\ref{fig6}(b) and \ref{fig6}(a2)-\ref{fig6}(f2)), the interference conditions between the bright rings keep changing, and as a result of intensity redistribution, the intensity on the rings becomes more nonuniform, as shown in Figs.\,\ref{fig6}(a,c), Figs.\,\ref{fig6}(a1)-\ref{fig6}(f1) and Figs.\,\ref{fig6}(a3)-\ref{fig6}(f3). The variation of k-value shows a similar trend, as it gradually increases from 1 to 10 until $z\approx300$ mm, and after that abruptly increases to a maximum value of $\sim 191$ at $z_{af}=387$ mm, and then drops abruptly and continues decreasing gradually. This again confirms an abruptly autofocusing behaviour of CADB in a turbulent medium with $\mathrm{S_R}=0.5$. However, the maximum k-value at $z_{af}=387$ ($\sim 191$ in simulation and $\sim 169$ in experiment) is found to be reduced as compared to the earlier cases of turbulent media with $\mathrm{S_R}=0.7$ and $\mathrm{S_R}=1$, indicating the reduced autofocusing efficiency. In particular, at $z_{af}$ in addition to a high-intensity peak some of the intensity also scattered in the background (Figs.\,\ref{fig6}(d1) and \ref{fig6}(d3)). Further, the autofocusing distance is also reduced slightly as compared to earlier cases of turbulent media with $\mathrm{S_R}=0.7$ and $\mathrm{S_R}=1$.
\begin{figure*}[htbp]
\centering
\includegraphics[height = 9.40cm, keepaspectratio = true]{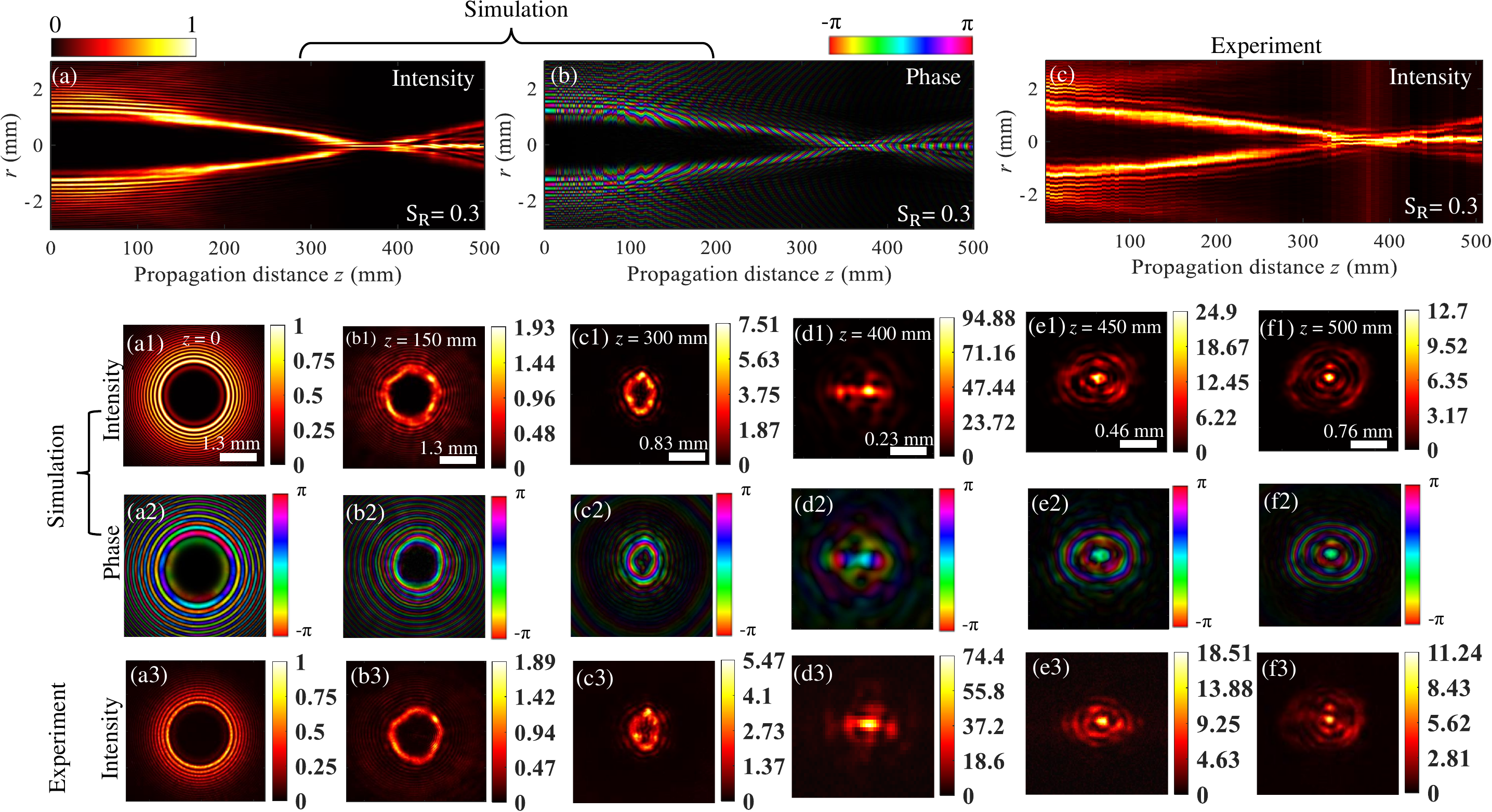}
\caption{In a turbulent media with $\mathrm{S_R}=0.3$, the intensity (a, c) and phase (b) distributions of a CADB as a function of propagation distance $z$. The snapshot of intensity and phase distributions of a CADB at propagation distance (a1, a2, a3) $z=0$, (b1, b2, b3) $z=150$ mm, (c1, c2, c3) $z=300$ mm, (d1, d2, d3) $z=400$ mm, (e1, e2, e3) $z=450$ mm, and (f1, f2, f3) $z=500$ mm in high turbulence. The other parameter values are taken as $r_0=1$ mm, $w_0=0.1$ mm, $a=0.1$ and $\lambda=1064$ nm.}
\label{fig7}
\end{figure*} 

Furthermore, we have propagated the CADB in a strong turbulent medium with $\mathrm{S_R}=0.3$, and analyzed the autofocusing abilities. The simulated and experimental results are shown in Fig.\,\ref{fig7}. Figures\,\ref{fig7}(a) and \ref{fig7}(b) show the simulated results of evolution of cross-sectional intensity and phase of CADB as a function of propagation distance. Figure\,\ref{fig7}(c) shows the experimental results of evolution of CADB's cross-sectional intensity as a function of propagation distance. Figures\,\ref{fig7}(a1)-\ref{fig7}(f1), Figs.\,\ref{fig7}(a2)-\ref{fig7}(f2) and Figs.\,\ref{fig7}(a3)-\ref{fig7}(f3) show the simulated intensity, simulated phase and experimental intensity distributions of CADB at different propagation distances, respectively. As evident, although turbulent media is significantly strong (Fig.\,\ref{fig2}(d)), the CADB still possesses a reasonably well autofocusing behaviour, which leads to the formation of a high-intensity peak at the autofocusing distance. Since a strong turbulent media consists of high randomness in the phase distribution, thus upon propagation the CADB experiences more prominent spatial distortions in its intensity distribution, as evidenced by the non-uniform intensity on the rings (Figs.\,\ref{fig7}(a), \ref{fig7}(c), \ref{fig7}(b1)-\ref{fig7}(f1) and \ref{fig7}(b3)-\ref{fig7}(f3)). The variation in the k-value still follow the same trend, as k-value gradually increases from 1 to $\sim 8$ until $z\approx300$ mm, and after that abruptly reaches to a maximum value of $\sim 95$, and then drops abruptly and continues decreasing gradually. This again confirms an abruptly autofocusing behaviour of CADB in a strong turbulent media with $\mathrm{S_R}=0.3$. However, the autofocusing efficiency is found to be reduced compared to earlier cases, as the maximum k-value reaches to $\sim 95$ [$\sim 75$] in the simulation [experiment]. The autofocusing distance is also found to be increased ($z_{af}=400$ mm) due to the strong turbulent media (Figs.\,\ref{fig7}(d1) and \ref{fig7}(d3)).

\begin{figure}
\centering
\includegraphics[height = 6.0cm, keepaspectratio = true]{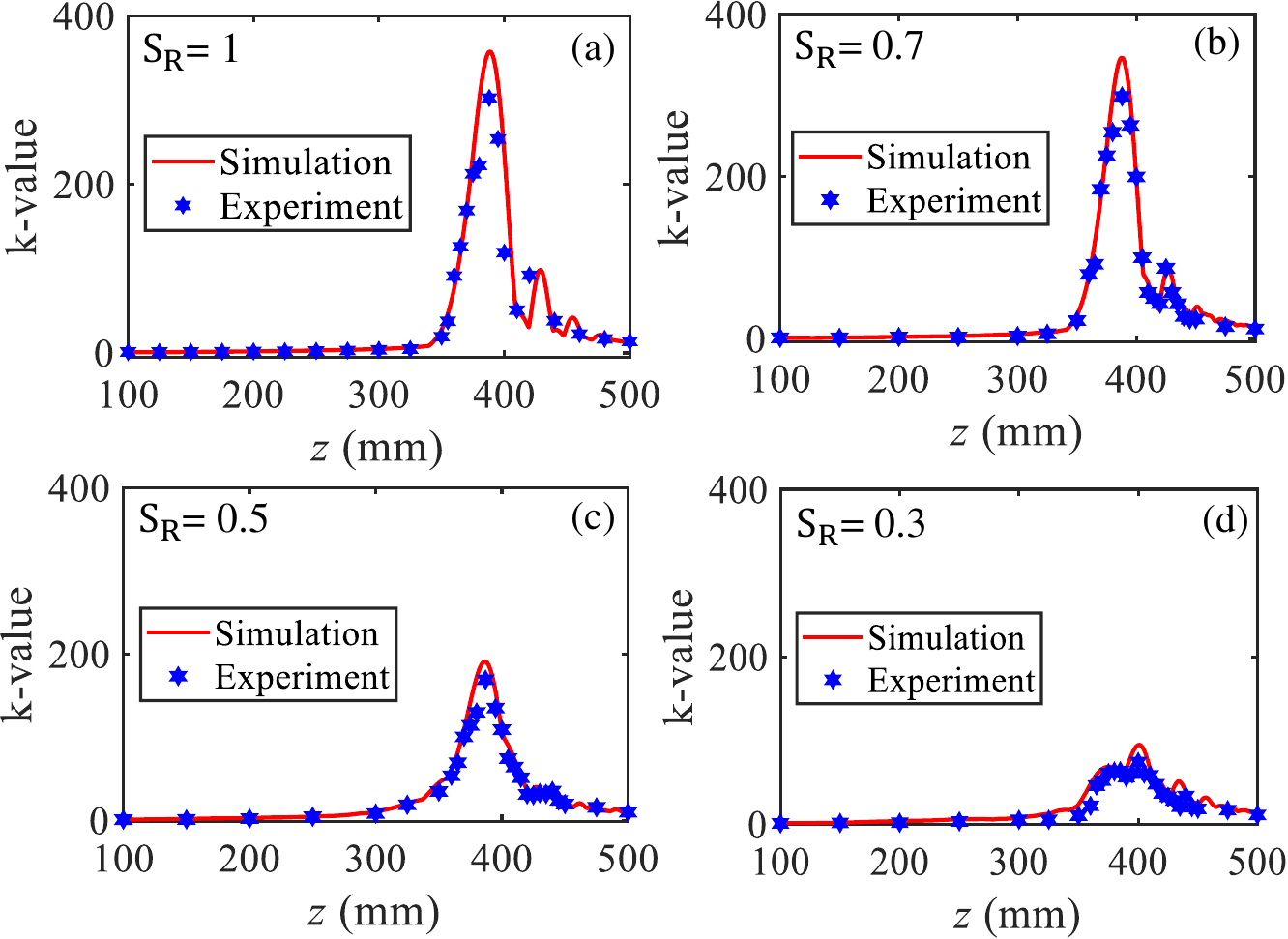}
\caption{The experimental (blue stars) and simulated (solid red curve) k-value of CADB as a function of propagation distance $z$ in (a) free space ($\mathrm{S_R}=1$), (b) low strength turbulence ($\mathrm{S_R}=0.7$), (c) medium strength turbulence ($\mathrm{S_R}= 0.5$), and (d) high strength turbulence ($\mathrm{S_R}=0.7$). The other parameter values are taken as $r_0=1$ mm, $w_0=0.1$ mm, $a=0.1$ and $\lambda=1064$ nm.}
\label{fig8}
\end{figure} 
A detailed quantified experimental and simulated results of k-value as a function of propagation distance $z$ in free space ($\mathrm{S_R}=1$) and turbulent media with different strengths ($\mathrm{S_R}=0.7$, 0.5 and 0.3) are shown in Figs.\,\ref{fig8}(a)-\ref{fig8}(d), respectively. The solid red curve and blue stars denote the simulation and experimental results, respectively. As evident, in all cases the k-value initially gradually increases with distance until $z\approx300$ mm, and after that it abruptly increases to a maximum value, and subsequently it drops rapidly, and then it continues to decrease gradually. Note, small oscillations in the k-value are observed after the autofocusing distance, which occurs due to multi-focusing behavior of CADB, as shown in Figs.\,\ref{fig4}(a) and \ref{fig5}(a). At these multi-focusing positions, the intensity is less tightly focused as compared at $z_{af}$. Furthermore, the maximum k-value is found to be highest in free space (Fig.\,\ref{fig8}(a)), and decreases with an increase of turbulence strength (Fig.\,\ref{fig8}(b)-\ref{fig8}(d)). These observations clearly suggest that abrupt autofocusing of CADB remains persisting in low to high strength turbulent media, however, the autofocusing efficiency decreases as evidenced by the reduced maximum k-value. The abruptness in autofocusing depends on how fast k-value changes just before and after $z_{af}$, and can be estimated by determining the full-width at half-maximum (FWHM) of k-value curve (Figs.\,\ref{fig8}(a)-\ref{fig8}(d)). Corresponding to the propagation of CADB in free space ($\mathrm{S_R}=1$) and turbulent media with $\mathrm{S_R}=0.7,~0.5$ and 0.3, the FWHMs of k-value curves are found to be 30 mm, 31 mm, 33 mm and 40 mm, respectively. It indicates that the abruptness of autofocusing has not affected significantly with increasing the turbulence strength. 

\begin{figure}
\centering
\includegraphics[height = 3.2cm, keepaspectratio = true]{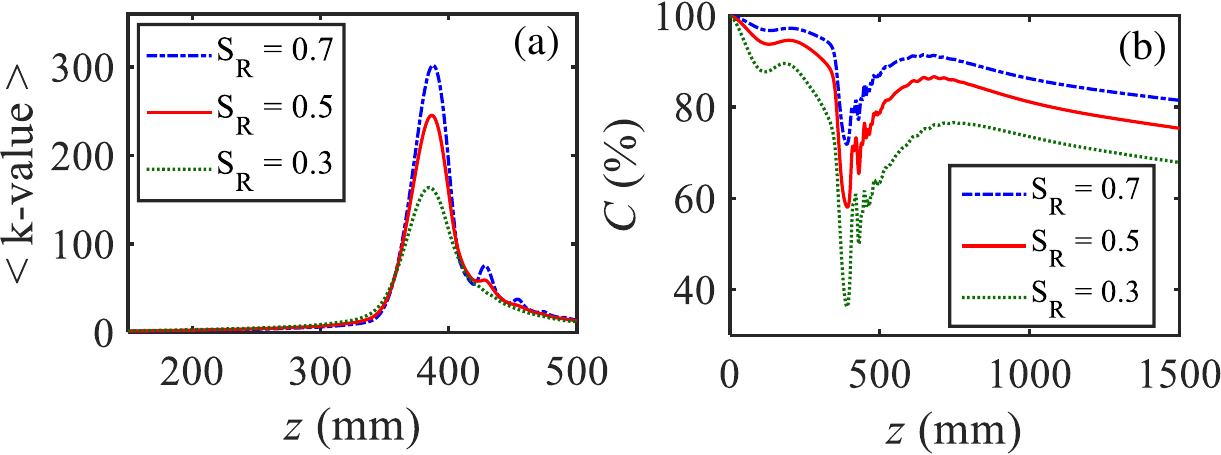}
\caption{Simulated results obtained by averaging over 80 realizations of turbulent media (random turbulent phase screens). (a) The average k-value as a function of propagation distance, for turbulent media with $\mathrm{S_R}= 0.7$ (dot-dashed blue curve), $\mathrm{S_R}= 0.5$ (solid red curve), and $\mathrm{S_R}=0.3$ (dotted green curve). (b) The overlap integral ($C$) as a function of propagation distance, for turbulent media with $\mathrm{S_R}= 0.7$ (dot-dashed blue curve), $\mathrm{S_R}= 0.5$ (solid red curve) and $\mathrm{S_R}= 0.3$ (dotted green curve).}
\label{fig9}
\end{figure}
So far, we have presented all the results for a fixed turbulent media having different strengths. The turbulent media of a desire length and fixed strength is realized by a phase screen constructed from a complex random matrix ($R_{NN}$) that can have several permutations. Further, in realistic situations the turbulent media can be dynamic in nature, where randomness can change locally, however, the overall turbulence strength remains the same. Therefore, for an statistical analysis, we have performed investigations over 80 realizations of different turbulence phase screens, and determine the maximum k-value, FWHM of k-value curve and autofocusing distance $z_{af}$. Note, the number of realizations more than 80 gives almost the same results. The detailed statistical analysis of results can be seen in Appendix\,\ref{appenA}. We have found that over different realizations of turbulent media, the values vary in a certain range, and particularly with increasing the strength of turbulent media, the range of variation gets enlarged. Therefore, for generalization of our study, we have calculated the average value over 80 different realizations of turbulent media with a given strength. 

The average k-value as a function of propagation distance for $\mathrm{S_R}=0.7$ (dot-dashed blue curve), $\mathrm{S_R}=0.5$ (solid red curve) and $\mathrm{S_R}=0.3$ (dotted green curve) is shown in Fig.\,\ref{fig9}(a). The averaged maximum k-value is found to be 300 (for $\mathrm{S_R}=0.7$), 246 (for $\mathrm{S_R}=0.5$), and 161 (for $\mathrm{S_R}=0.3$), indicating that the autofocusing efficiency of CADB in turbulent media decreases with the increase of strength of turbulence. However, the abrupt autofocusing is found in all three cases. The FWHMs of average k-value curves are found to be 31.5 mm (for $\mathrm{S_R}=0.7$), 34.2 mm (for $\mathrm{S_R}=0.5$), and 42.2 mm (for $\mathrm{S_R}=0.7$), indicating that the abruptness of autofocusing has also slightly decreased with increasing the turbulence strength. Further, the average value of $z_{af}$ is found to be 388 mm, 387 mm, and 385 mm for $\mathrm{S_R} = 0.7$, $\mathrm{S_R}=0.5$, and $\mathrm{S_R}=0.3$ respectively (see Appendix\,\ref{appenA}). 

The propagation of CADB in turbulent media leads to spatial intensity distortions as shown in Figs.\,\ref{fig5}-\ref{fig7}. To quantify the spatial intensity distortions in CADB induced by turbulent media, we calculate an overlap integral, which is given by \cite{dev2021}
\begin{equation}
C(z)=\frac{\int \int I_f(x,y;z)I_t(x,y;z)dxdy}{\sqrt{\int \int I_f^{2}(x,y;z)dxdy}\sqrt{\int \int I_t^{2}(x,y;z)dxdy}}, \label{eq15}
\end{equation}
where $I_f(x,y;z)$ and $I_t(x,y;z)$ denote the intensities of CADB in free space and turbulent media, respectively. The calculated overlap integral as a function of propagation distance, for different turbulence strength $\mathrm{S_R=0.7}$ (dot-dashed blue curve), $\mathrm{S_R=0.5}$ (solid red curve) and $\mathrm{S_R=0.3}$ (dotted green curve) is shown in Fig.\,\ref{fig9}(b). Note, the overlap integral is averaged over 80 realizations of turbulent media. As evident, the overlap integral initially decreases slowly, and close to $z_{af}$ it drops rapidly and becomes minimum at $z_{af}$, and after that it again increases and acquires almost a constant value. For turbulent media with $\mathrm{S_R=0.7,~0.5}$ and 0.3, at $z_{af}$ the minimum values of $C$ are found to be $71\%$, $58\%$ and $36\%$, respectively. The overlap integral remains reasonably well except around $z_{af}$. This is due to the fact that at $z_{af}$ the entire CADB intensity is tightly focused in a small area, and it remains confined at values near to $z_{af}$, therefore, a small deviation in CADB (spatial distortion and beam wandering) due to turbulent media leads to a significant change in $C$. Further, the overlap $C$ decreases with increasing the strength of turbulence. However, it remains reasonably well, indicating the turbulence resilience characteristics of CADBs.

\begin{figure}
\centering
\includegraphics[height = 6.4cm, keepaspectratio = true]{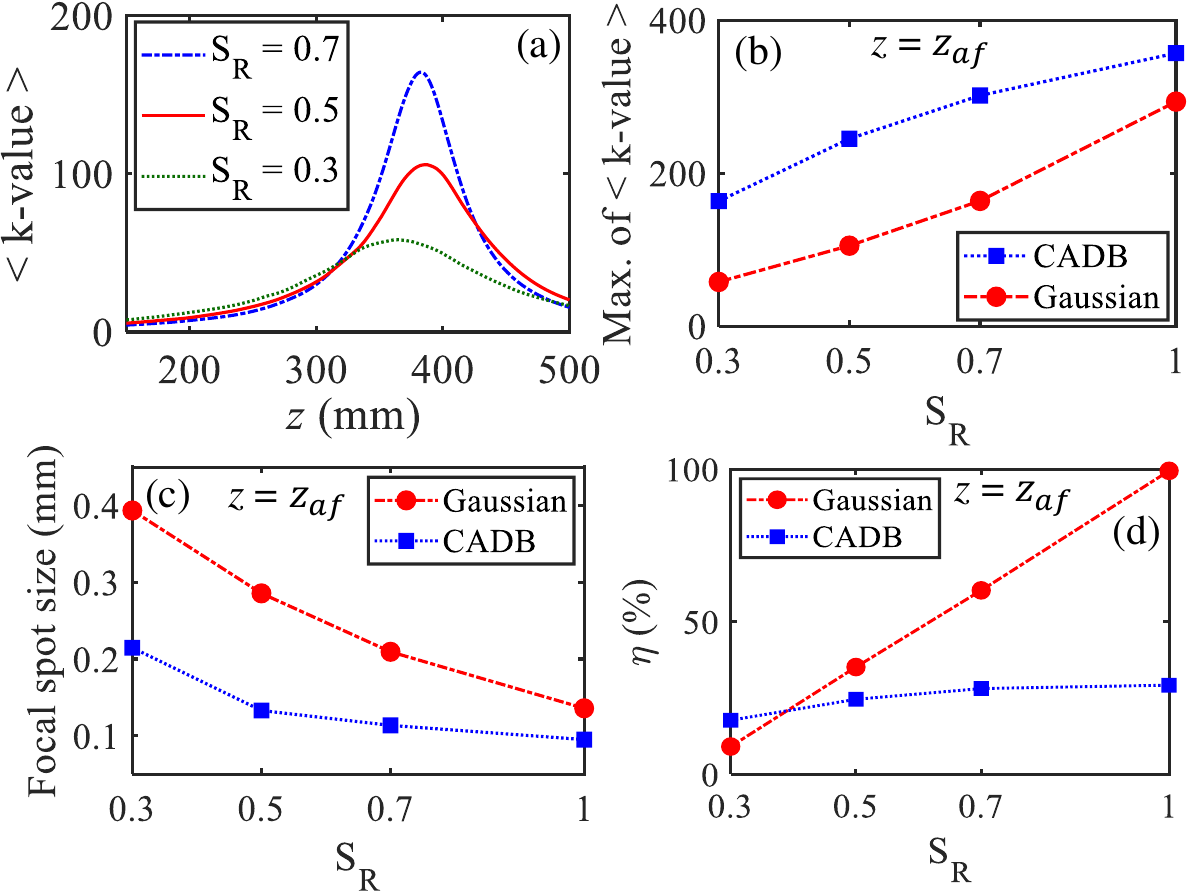}
\caption{Simulated results obtained by averaging over 80 realizations of turbulent media (random turbulent phase screens). (a) The average k-value of Gaussian beam as a function of $z$, for turbulent media with $\mathrm{S_R}= 0.7$ (dot-dashed blue curve), $\mathrm{S_R}= 0.5$ (solid red curve), and $\mathrm{S_R}=0.3$ (dotted green curve). (b) At $z_{af}$, the average k-value of CADB and Gaussian as a function of turbulence strength $\mathrm{S_R}$. (c) At $z_{af}$, the spot size of CADB and Gaussian as a function turbulence strength $\mathrm{S_R}$. (d) At $z_{af}$, the diffraction efficiency of CADB and Gaussian as a function of turbulence strength ($\mathrm{S_R}$). The blue filled squares and red filled circles denote the results for CADB and Gaussian beam, respectively.}
\label{fig10}
\end{figure}

Earlier investigations show that structured light may possess better propagation properties in complex media than a conventional Gaussian beam \cite{Cox2021,Forbes2021,dev2021}. Therefore, to show the advantages we have compared the results of CADB with a conventional Gaussian beam under similar conditions. To do that we have investigated the propagation of a Gaussian beam with and without a focusing lens in free space and turbulent media with $\mathrm{S_{R}}=0.7,~0.5$ and 0.3 (see Appendix\,\ref{appenB}). Note, for the propagation analysis of a Gaussian beam in turbulent media with different strengths, we have considered a single realization of turbulent media as it was considered for CADB in Figs.\,\ref{fig5}-\ref{fig7}. It is shown that propagation of a Gaussian beam, without a focusing lens, in free space leads to increase in the beam size due to diffraction. However, in turbulent media, the Gaussian beam experiences strong distortions and leads to speckled intensity distribution (see Fig.\,\ref{fig21} in Appendix\,\ref{appenB}). Upon increasing the strength of turbulent media the speckles become more pronounced, which indicates that Gaussian beam is highly sensitive to the turbulence (Appendix\,\ref{appenB}). 

The propagation of a Gaussian beam, in free space, through a plano-convex lens of focal length $f=388$ mm (equivalent to autofocusing distance of CADB) results a tightly focused beam at a distance $z=388$ mm. However, in turbulent media, the propagation of Gaussian beam results strong spatial distortions (increases with increasing strength of turbulence), and focusing ability degrades significantly (see Fig.\,\ref{fig22} in Appendix\,\ref{appenB}). It indicates that the Gaussian beam pose strong limitations in complex media. 

For the generalization, we have again performed investigations over 80 realizations of turbulent media, and obtained the averaged k-value. The average k-value of a Gaussian beam as a function of propagation distance in turbulent media with different strength is shown in Fig.\,\ref{fig10}(a). Particularly, the results are shown for turbulent media with $\mathrm{S_{R}}=0.7$ (dot-dashed blue curve), $\mathrm{S_{R}}=0.5$ (solid red curve), and $\mathrm{S_{R}}=0.3$ (dotted green curve). As evident, the efficiency of focusing of a Gaussian beam reduces considerably with the increase of strength of turbulence. Further, unlike CADB (Fig.\,\ref{fig9}(a)), the Gaussian beam does not show good abrupt focusing in turbulent media.

A detailed comparison between CADB and Gaussian beam is presented in Figs.\,\ref{fig10}(b)-\ref{fig10}(d), where at the focusing distance $z_{af}$, the maximum $\langle$k-value$\rangle$, focal spot-size, and diffraction efficiency ($\eta$) are compared for different strength of turbulence. These results are obtained by averaging over 80 realizations of turbulent media (turbulent phase screens). For both Gaussian and CADB, the maximum $\langle$k-value$\rangle$ decreases with increasing the strength of turbulence, however, it remains higher for CADB (Fig.\,\ref{fig10}(b)) that indicates better-focusing efficiency in turbulent media. Further, the focal spot size refers to the $1/e^2$ width which is equal to the distance between the two points on the intensity distribution of CADB/Gaussian beam at $z_{af}$ where intensity becomes $1/e^2 = 0.135$ times of maximum value. As evident, the focal spot size increases more significantly for a Gaussian beam (Fig.\,\ref{fig10}(c)). Specifically, for a high strength turbulent media with $\mathrm{S_R}=0.3$, the spot size grows by a factor of $2.9$ for a Gaussian beam and $2.2$ for a CADB, indicating that CADB possesses better focusing abilities in complex media. Figure\,\ref{fig10}(d) shows the diffraction efficiency of Gaussian beam (red filled circles) and CADB (blue filled squares), as a function of strength of turbulence. The diffraction efficiency is calculated by taking ratio between the focused beam intensity (intensity within the main central peak) at $z=z_{af}$ and initial beam intensity at $z=0$ \cite{dev2022}. As evident, for a Gaussian beam the diffraction efficiency decreases significantly as a function of strength of turbulent media, in particular it changes by a factor of $10.8$ for varying media with $\mathrm{S_{R}}=1$ to $\mathrm{S_{R}}=0.3$. Whereas, for CADB the diffraction efficiency changes by a factor of 1.7, for the same variation of turbulence strength. It clearly suggests that CADB possesses better resilience and ability to focus in strong turbulent media. 
\section{Explanation of results}
\label{Explan}
\begin{figure*}[htbp]
\centering
\includegraphics[height = 3.8cm, keepaspectratio = true]{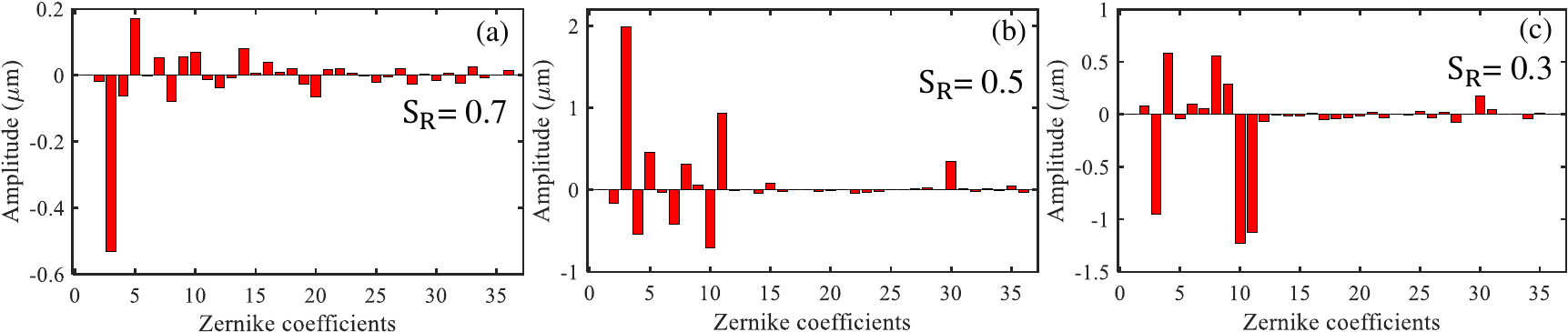}
\caption{The amplitude of Zernike coefficients present in turbulent media with varying strength (a) $\mathrm{S_R}=0.7$ (low), (b) $\mathrm{S_R}=0.5$ (medium), and (c) $\mathrm{S_R}=0.3$ (high). Note, the analysis is presented for single realizations of turbulent media with different $\mathrm{S_R}$ (Fig.\,\ref{fig2}), and are used in Figs.\,\ref{fig5}-\ref{fig7}.}
\label{fig11}
\end{figure*}
As discussed above, when CADB propagates in turbulent media with varying strength several effects occurs, such as, spatial distortions, change in $z_{af}$, change in k-value, change in abruptness of autofocusing, change in transverse position of autofocusing, and change in the diffraction efficiency. To explain these effects, we decompose the turbulent media into a series of Zernike polynomials \cite{burger2008,noll1976,mphuthi2018}, which are a set of mathematical functions often used to describe the phase aberrations or distortions in optical systems, including those induced by turbulence. These polynomials provide a convenient and efficient way to express complex wavefront distortions, allowing us to analyze and correct for the effects of turbulence in various optical applications. They form an orthogonal basis set over the unit circle and are scaled to unit variance.
The polynomials are defined as \cite{noll1976,mphuthi2018,niu2022}
\begin{equation}
Z_{p,q}(r,\theta)=\left\{\begin{array}{cl}
 U_p^q(r,\theta) :& q < 0; |q-p|=\mathrm{even},\\
 V_p^q(r,\theta) :&  q \neq 0;  |q-p|=\mathrm{odd},\\
 \sqrt{(p+1)} R_p^0(r) :&  q = 0, \\
\end{array}\right. 
\label{eq16}
\end{equation}
where  
\begin{eqnarray}
 U_p^q(r,\theta) &=& \sqrt{2(p+1)}~R_p^q(r)\cos(q\theta), \nonumber\\
 V_p^q(r,\theta) &=& \sqrt{2(p+1)}~R_p^q(r)\sin(q\theta), \nonumber
 \end{eqnarray}
and 
 \begin{equation}
R_p^q(r) = \sum_{s=0}^{\frac{(p-q)}{2}}\frac{(-1)^s(p-s)!}{s(\frac{p+q}{2}-s)!(\frac{p-q}{2}-s)!}r^{p-2s}.
 \nonumber
 \end{equation}
Here, $p$ and $q$ are the radial and azimuthal indices, which are integers and follow the Noll convention \cite{niu2022}. The orthogonality of Zernike basis functions enables us to represent any turbulent media as a sum of weighted Zernike polynomials, which is given by
 \begin{equation}
 \psi'(r,\theta) = \sum_{p=0}^{\infty}\sum_{q=0}^{p}a_{p,q}~Z_{p,q}(r,\theta),
 \end{equation}
where polynomial $Z_{pq}$ and coefficient $a_{pq}$ represent the type and amplitude of aberration present, respectively. To explain the turbulence induced effects on the properties of CADB, we initially decompose the turbulent media into a combination of Zernike polynomials up to $36^{\mathrm{th}}$ orders. Subsequently, we determine the type and amplitude of each aberration in units of wavelength. Figures\,\ref{fig11}(a), \ref{fig11}(b), and \ref{fig11}(c) show the calculated amplitude of aberrations present in turbulent media with strength $\mathrm{S_R}=0.7$ (Fig.\,\ref{fig2}(b)), $\mathrm{S_R}=0.5$ (Fig.\,\ref{fig2}(c)), and $\mathrm{S_R}=0.3$ (Fig.\,\ref{fig2}(c)), respectively. Note, here we have considered a single realization of turbulent media with different turbulent strength. As evident, in all three different turbulent media, the Zernike coefficients consist of different amplitudes, indicating the dominance of different aberrations. Following this, our investigation involves analysis of the impact of each aberration that is present in a quantity sufficient to induce an undesired effect on the beam. To achieve this, we create a phase screen with only a single aberration at the calculated magnitude ($a_{p,q}$), and we add this phase fluctuation in the beam, propagate it, and then measure the effect on the beam. 
\subsection{y-tilt aberration}
The $x$/$y$-tilt aberration gives rise to the effect of beam wandering which refers to the deviation of the beam's path from its intended trajectory. As shown in Fig.\,\ref{fig11}, the amplitude of $x$-tilt aberration is found to be very small having a negligible effect on the beam's trajectory as compared to $y$-tilt aberration which is represented by the $3^{\mathrm{rd}}$ order Zernike polynomial ($Z_{1,-1}$) as \cite{niu2022}
 \begin{equation}
 Z_{1,-1}(r,\theta) = 2r\sin\theta.
 \end{equation}
\begin{figure}[htbp]
\centering
\includegraphics[height = 3.70cm, keepaspectratio = true]{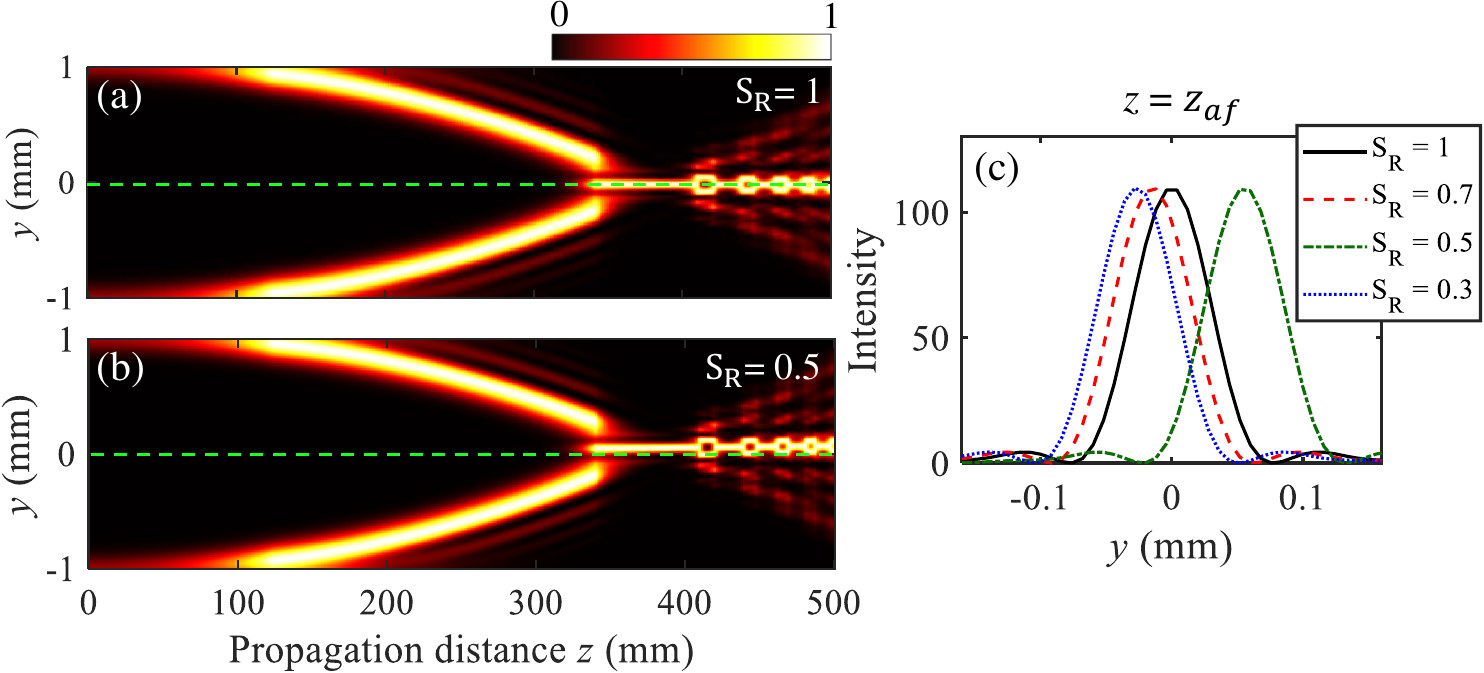}
\caption{The cross-sectional intensity distribution of CADB as a function of propagation distance in turbulent media with (a) $\mathrm{S_R}=1$, and (b) $\mathrm{S_R}=0.5$. (c) Intensity cross-sections taken along the y-axis in (a) $\&$ (b) at $z=z_{af}$. Note, here turbulent media corresponds to a phase screen containing only tilt aberration with certain amplitude.}
\label{fig12}
\end{figure}
The amplitudes of $Z_{1,-1}$ are found to be $-0.53 ~\mu$m, $1.98 ~\mu$m, and $-0.95 ~\mu$m in turbulent media with $\mathrm{S_R}=0.7$, $\mathrm{S_R}=0.5$, and $\mathrm{S_R}=0.3$, respectively (Fig.\,\ref{fig11}). Corresponding to these amplitudes we create turbulent phase screens containing only this aberration, and then impose these phase screens onto CADB (separately for each case), and after propagation of CADB, the effects are quantified. The results are shown in Fig.\,\ref{fig12}.
Figures\,\ref{fig12}(a) and \ref{fig12}(b) show the cross-sectional intensity distribution of CADB as a function of propagation distance in free space ($\mathrm{S_R}=1$) and turbulent media with $\mathrm{S_R}=0.5$, respectively. The dashed green line marks the on-axis position ($y=0$). Figure\,\ref{fig12}(c) shows the intensity cross-section taken along the $y-$axis at $z=z_{af}$ in Figs.\ref{fig12}(a) and \ref{fig12}(b). Note, the cross-sectional intensity distribution of CADB for $\mathrm{S_R}=0.7$ and $\mathrm{S_R}=0.3$ are not shown. As evident, in free space ($\mathrm{S_R}=1$), the CADB shows the autofocusing at on-axis position (Fig.\,\ref{fig12}(a)). However, in turbulent media with different strength, the CADB undergoes beam wandering that leads to the autofocusing at off-axis position (intensity shifted vertically with respect to the dashed green line in Fig.\,\ref{fig12}(b)). Further, the beam wandering occurs either in +/$-$ $y$-direction depending on the sign of amplitude of y-tilt aberration. For example, in turbulent media with $\mathrm{S_R}=0.5$ the amplitude of aberration is $1.98 ~\mu$m (with positive sign), which causes the beam shifting of 0.05 mm in the $+y$ direction (Fig.\,\ref{fig12}(c)). In contrast, for $\mathrm{S_R}=0.7$ and $0.3$ the amplitudes are $-0.53 ~\mu$m and $-0.95 ~\mu$m (with negative sign), leading to the beam shifting of 0.01 mm and 0.02 mm in negative y-direction, respectively (Fig.\,\ref{fig12}(c)).
\subsection{Defocus aberration}
Defocus aberration refers to the impact of varying refractive index gradients in a media on the focus of optical systems that can result in blurring and distortions of images. Defocus aberration is represented by $4^{\mathrm{th}}$ order Zernike polynomial ($Z_{2,0}$), which is given as \cite{niu2022}
 \begin{equation}
 Z_{2,0}(r,\theta) = \sqrt{3}(2r^2-1).
 \end{equation}
\begin{figure}[htbp]
\centering
\includegraphics[height = 3.70cm, keepaspectratio = true]{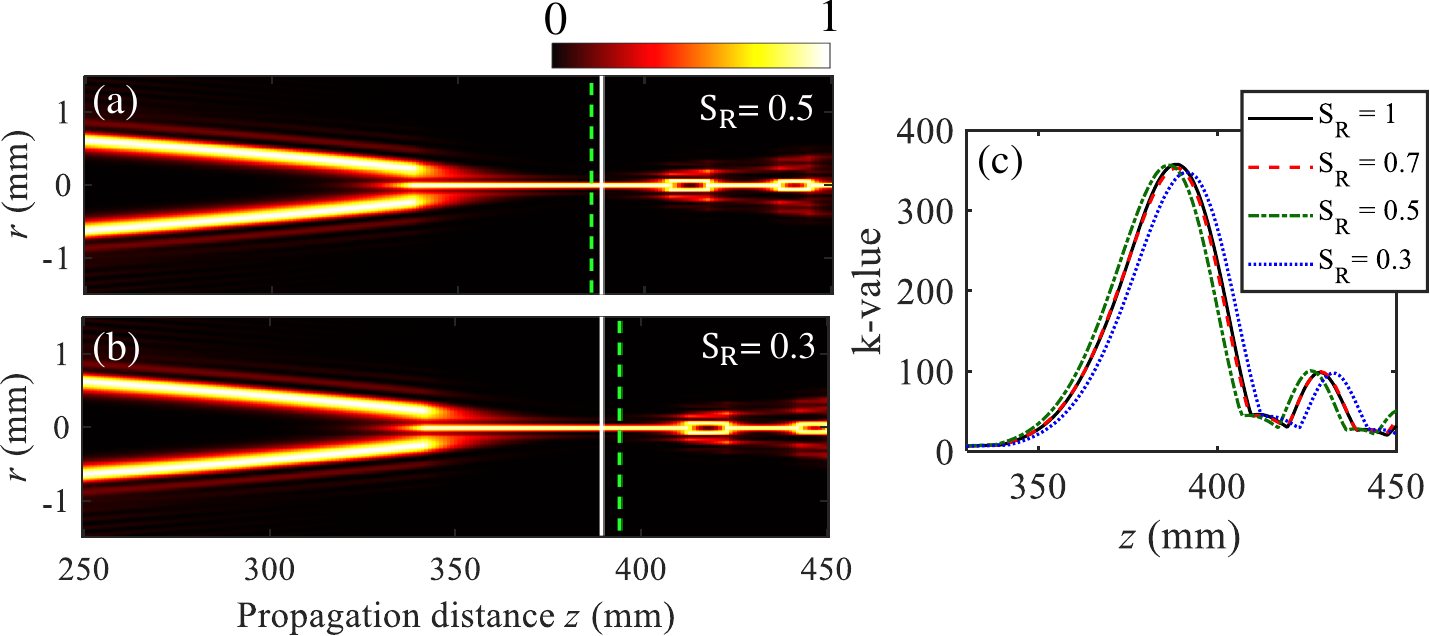}
\caption{The cross-sectional intensity distribution of CADB as a function of propagation distance in turbulent media with (a) $\mathrm{S_R}=0.5$, and (b) $\mathrm{S_R}=0.3$. Note, here turbulent media corresponds to a phase screen containing only defocus aberration with certain amplitude. The white solid line and dashed green line mark the location of $z_{af}$ in free space ($\mathrm{S_R}=1$) and in turbulent media ($\mathrm{S_R}=0.5$ and $\mathrm{S_R}=0.3$), respectively. (c) The k-value as a function of propagation distance for free space ($\mathrm{S_R}=1$) (solid black curve), and turbulent media with $\mathrm{S_R}=0.7$ (dashed red curve),  $\mathrm{S_R}=0.5$ (dot-dashed green curve) and $\mathrm{S_R}=0.3$ (dotted blue curve).}
\label{fig13}
\end{figure}

The amplitudes of $Z_{2,0}$ are found to be $-0.06 ~\mu$m, $-0.53 ~\mu$m, and $0.58 ~\mu$m in turbulent media with $\mathrm{S_R}=0.7$, $\mathrm{S_R}=0.5$, and $\mathrm{S_R}=0.3$, respectively (Fig.\,\ref{fig11}). We have again adapted the same procedure as earlier, and analyzed the effect of this aberration on the propagation of CADB, as shown in Fig.\,\ref{fig13}. Figures\,\ref{fig13}(a) and \ref{fig13}(b) show the cross-sectional intensity distribution of CADB as a function of propagation distance for turbulent media with $\mathrm{S_R}=0.5$ and $\mathrm{S_R}=0.3$, respectively. Note, white solid line and dashed green line mark the location of $z_{af}$ in free space ($\mathrm{S_R}=1$) and in turbulent media (with $\mathrm{S_R}=0.5$ and $\mathrm{S_R}=0.3$). Figure\,\ref{fig13}(c) shows the k-value as a function of propagation distance for free space ($\mathrm{S_R}=1$) (solid black curve) and turbulent media with $\mathrm{S_R}=0.7$ (dashed red curve),  $\mathrm{S_R}=0.5$ (dot-dashed green curve) and $\mathrm{S_R}=0.3$ (dotted blue curve). As evident, corresponding to the amplitudes $a_{2,0}= -0.06 ~\mu$m (for $\mathrm{S_R}=0.7$), $a_{2,0}=-0.53 ~\mu$m (for $\mathrm{S_R}=0.5$), and $a_{2,0}=0.58 ~\mu$m (for $\mathrm{S_R}=0.3$), the values of $z_{af}$ are found to be 388 mm, 386 mm, 392 mm.  A decrease [increase] in $z_{af}$ depends on the negative [positive] sign of amplitude of defocus aberration. It clearly indicates that the presence of defocus aberration in turbulent media changes the autofocusing distance of CADB.
\subsection{Oblique Astigmatism aberration}
In oblique astigmatism aberration, light rays from a point source, when passing through an optical system, do not converge to a single focal point. Instead, these focus at different axial positions, resulting in the formation of a pair of line images \cite{liu2016}. This aberration is represented by $5^{\mathrm{th}}$ order Zernike polynomial ($Z_{2,-2}$), which is given as
\begin{equation}
 Z_{2,-2}(r,\theta) = \sqrt{6}(r^2\sin2\theta).
 \end{equation}
 
\begin{figure}[htbp]
\centering
\includegraphics[height = 4.0cm, keepaspectratio = true]{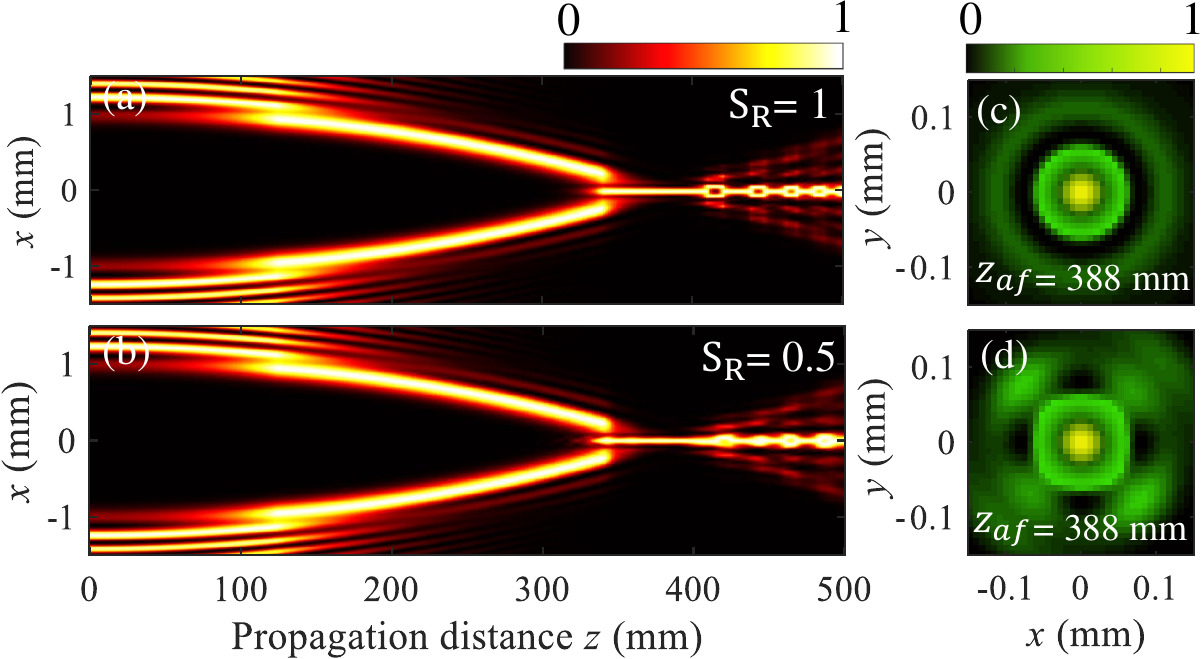}
\caption{The cross-sectional intensity distribution of CADB as a function of propagation distance in (a) free space ($\mathrm{S_R}=1$), and (b) turbulent media with $\mathrm{S_R}=0.5$. Note, here turbulent media corresponds to a phase screen containing only oblique astigmatism aberration with certain amplitude. The intensity distribution of CADB at $z_{af}=388$ mm in (c) free space ($\mathrm{S_R}=1$), and (d) turbulent media with $\mathrm{S_R}=0.5$. Note, for better visualization, we have chosen a different color code in (c) and (d).}
\label{fig14}
\end{figure}

Corresponding to the turbulent media with $\mathrm{S_R}= 0.7, ~0.5,~ \mathrm{and}~0.3$, the amplitudes of $Z_{2,-2}$ are found to be $0.17 ~\mu$m, $0.45 ~\mu$m, and $-0.04 ~\mu$m, respectively. The effect of this aberration with the largest amplitude $a_{2,-2}=0.45 ~\mu$m on the propagation of CADB is shown Fig.\,\ref{fig14}. Figures\,\ref{fig14}(a) and \ref{fig14}(b) show the cross-sectional intensity distribution of CADB as a function of propagation distance for free space ($\mathrm{S_R}=1$) and turbulent media with $\mathrm{S_R}=0.5$, respectively. Figures\,\ref{fig14}(c) and \ref{fig14}(d) show the intensity distribution of CADB at $z_{af}=388$ mm for free space ($\mathrm{S_R}=1$) and turbulent media with $\mathrm{S_R}=0.5$, respectively. As evident, the presence of this aberration caused distortion in the intensity distribution of CADB at $z_{af}$, and made it radially asymmetric (Fig.\,\ref{fig14}(d)) in comparison to the case of $\mathrm{S_{R}}=1$ (Fig.\,\ref{fig14}(c)). For better visualization, we have chosen a different color code in Figs.\,\ref{fig14}(c) and \ref{fig14}(d).
\subsection{Coma aberration} \label{comaabr}
Coma aberration occurs when off-axis light rays do not converge to a single point, resulting in distorted image of a point source. The vertical [horizontal] coma aberration affects the image in vertical [horizontal] direction. The turbulent media, characterized by fluctuations in the refractive index in terms of different temperatures, can cause light rays to bend and shift unpredictably as they travel through the medium. This bending and shifting can introduce vertical/horizontal coma aberrations, leading to distorted and elongated shapes. 
\begin{figure}[htbp]
\centering
\includegraphics[height = 4.20cm, keepaspectratio = true]{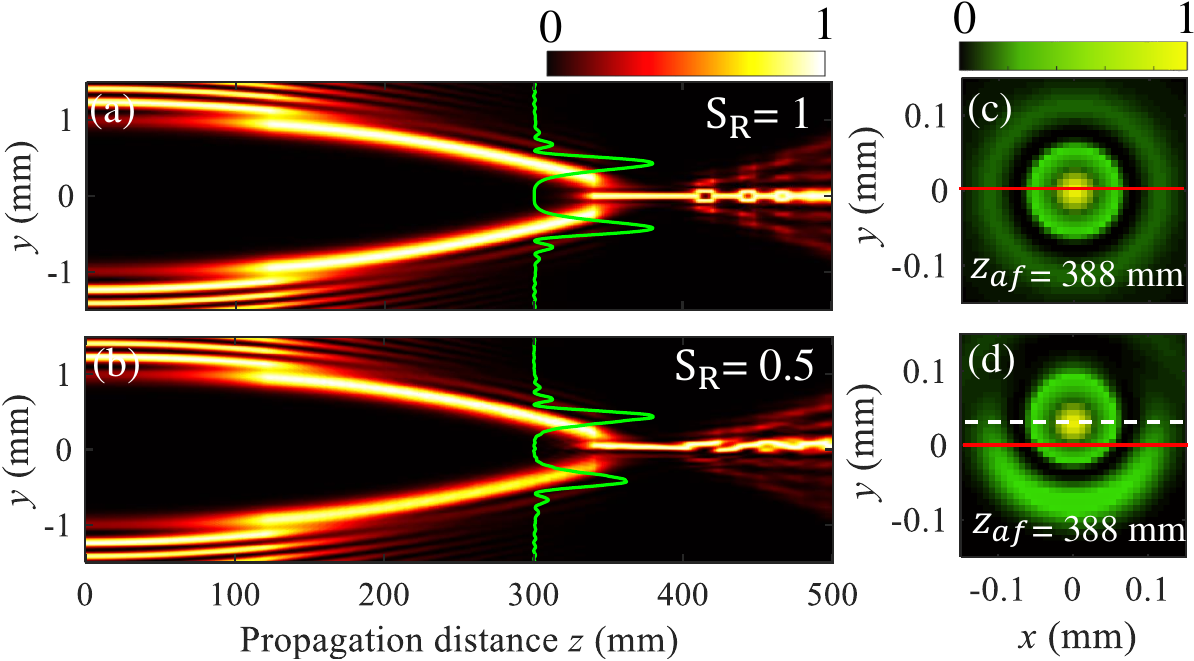}
\caption{Effect of vertical Coma aberration. The cross-sectional intensity distribution of CADB in (a) free space ($\mathrm{S_R}=1$), and (b) turbulent media with $\mathrm{S_R}=0.5$. The green solid curve represents the intensity cross-section at $z=300$ mm. Note, here turbulent media corresponds to a phase screen containing only vertical coma aberration with certain amplitude.The intensity distribution of CADB at $z_{af}=388$ mm in (c) free space ($\mathrm{S_R}=1$), and (d) turbulent media with $\mathrm{S_R}=0.5$. The red solid line and dashed white line mark the vertical position of focused beam spot for without aberration ($\mathrm{S_R}=1$) and with aberration ($\mathrm{S_R}=0.5$), respectively.}
\label{fig15}
\end{figure}

The vertical Coma aberration is represented by $7^{\mathrm{th}}$ order Zernike polynomial, which is given as \cite{niu2022}
\begin{equation}
 Z_{3,-1}(r,\theta) = \sqrt{8}(3r^3-2r)\sin\theta.
 \end{equation} 
 In the turbulent media with $\mathrm{S_R}=0.7$, $\mathrm{S_R}=0.5$, and $\mathrm{S_R}=0.3$, the amplitudes of $Z_{3,-1}$ are found to be $0.05 ~\mu$m, $-0.41 ~\mu$m, and $0.05 ~\mu$m, respectively. Since the amplitude of vertical coma aberration is largest in the case of $\mathrm{S_R}=0.5$, so we have analyzed the effect of vertical coma aberration on the propagation of CADB for only this case, as shown in Fig.\,\ref{fig15}. For comparison, we have also included results for $\mathrm{S_{R}}=1$. Figures\,\ref{fig15}(a) and \ref{fig15}(b) show the cross-sectional intensity distribution of CADB as a function of propagation distance for free space ($\mathrm{S_R}=1$) and turbulent media with $\mathrm{S_R}=0.5$, respectively. Figures\,\ref{fig15}(c) and \ref{fig15}(d) show the intensity distribution of CADB at $z_{af}=388$ mm in free space ($\mathrm{S_R}=1$) and turbulent media with $\mathrm{S_R}=0.5$, respectively. The solid red line marks the vertical position of focused beam spot when no aberration is present ($\mathrm{S_R}=1$). As evident, in turbulent media with $\mathrm{S_R}=0.5$, the focused beam spot (high-intensity peak) has been shifted vertically along the y-axis (shown by the dashed white line in Fig.\,\ref{fig15}(d)). Further, the intensity has also become asymmetric along the y-axis (Figs.\,\ref{fig15}(b) and \ref{fig15}(d)), as compared to the case for $\mathrm{S_{R}}=1$ ((Figs.\,\ref{fig15}(a) and \ref{fig15}(c)). 

\begin{figure}[htbp]
\centering
\includegraphics[height = 4.2cm, keepaspectratio = true]{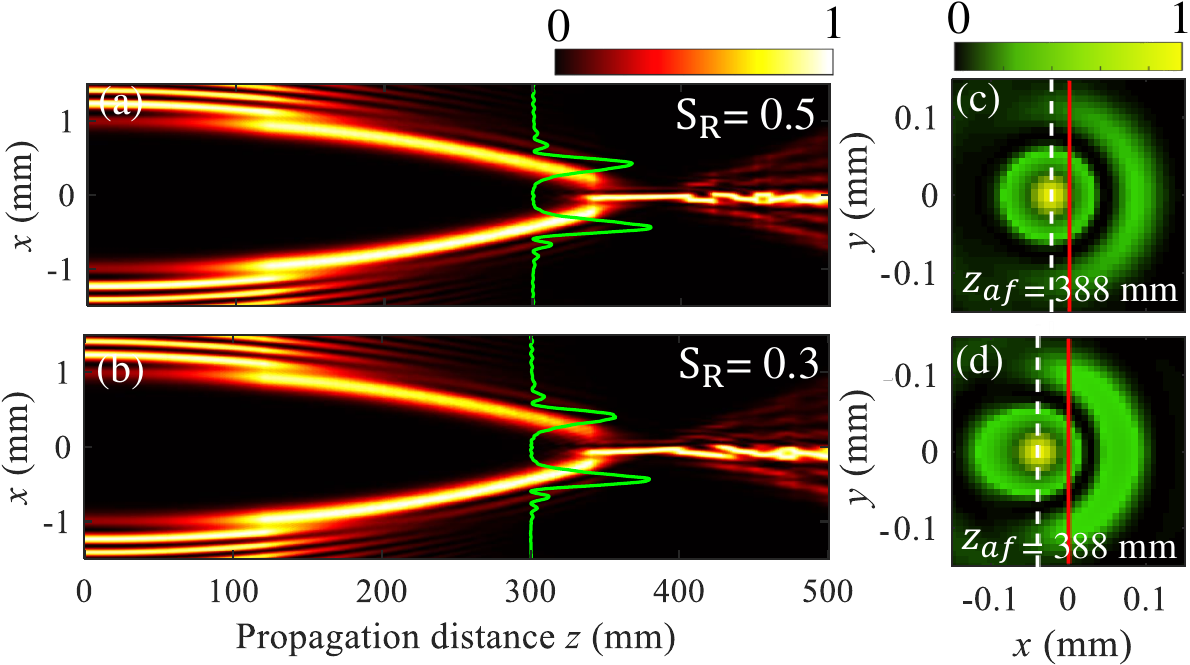}
\caption{Effect of horizontal Coma aberration. The cross-sectional intensity distribution of CADB in turbulent media with (a) $\mathrm{S_R}=0.5$, and (b) $\mathrm{S_R}=0.3$. The solid green curve represents the intensity cross-section at $z=300$ mm. Note, here turbulent media corresponds to a phase screen containing only horizontal coma aberration with certain amplitude. The intensity distribution of CADB at $z_{af}=388$ mm in turbulent media with (a) $\mathrm{S_R}=0.5$, and (b) $\mathrm{S_R}=0.3$. The red solid line and dashed white line mark the horizontal position of focused beam spot for without aberration ($\mathrm{S_R}=1$) and with aberration ($\mathrm{S_R}=0.5$ and $\mathrm{S_R}=0.3$), respectively.}
\label{fig16}
\end{figure}

 The similar effects on the CADB can also occur along the horizontal direction (x-axis), which are caused by the horizontal Coma aberration. It is represented by the $8^{\mathrm{th}}$ order Zernike polynomial ($Z_{3,1}$), which is given as \cite{niu2022}
 \begin{equation}
 Z_{3,1}(r,\theta) = \sqrt{8}(3r^3-2r)\cos\theta.
 \end{equation} 
In the turbulent media with $\mathrm{S_R}=0.7$, $\mathrm{S_R}=0.5$, and $\mathrm{S_R}=0.3$, the amplitudes of $Z_{3,1}$ are found to be $-0.07 ~\mu$m, $0.31 ~\mu$m, and $0.56 ~\mu$m, respectively. The results for horizontal Coma aberration induced effects on the CADB are shown in Fig.\,\ref{fig16}. Note, the amplitude corresponding to $\mathrm{S_R}=0.7$ is very small, so it is not shown. As evident, due to horizontal Coma aberration, the focused beam spot at $z_{af}$ (marked by dashed white lines in Figs.\,\ref{fig16}(c) and \ref{fig16}(d)) has been shifted with respect to central position $x=0$ (marked by solid red line, position of focused beam spot when no aberration is present) by an amount corresponding to the amplitudes of horizontal Coma aberration. Further, the horizontal Coma aberration also leads to asymmetry in intensity distribution of CADB along the x-axis.

It can be clearly seen that Coma aberration leads to the beam wandering as well as asymmetry in the intensity distribution of CADB.
\subsection{Vertical Trefoil aberration}
The vertical trefoil aberration is an optical distortion characterized by the presence of threefold symmetry in the wavefront of light, especially impacting the optical field along the vertical (y-axis) direction. This aberration can cause the focal point of an optical system to form a distorted trefoil-shaped pattern rather than a precise point. This aberration is presented by $9^{\mathrm{th}}$ order Zernike polynomial ($Z_{3,-3}$), which is given as \cite{niu2022}
\begin{equation}
 Z_{3,-3}(r,\theta) = \sqrt{8}(r^3\sin3\theta).
 \end{equation} 
\begin{figure}[htbp]
\centering
\includegraphics[height = 2.80cm, keepaspectratio = true]{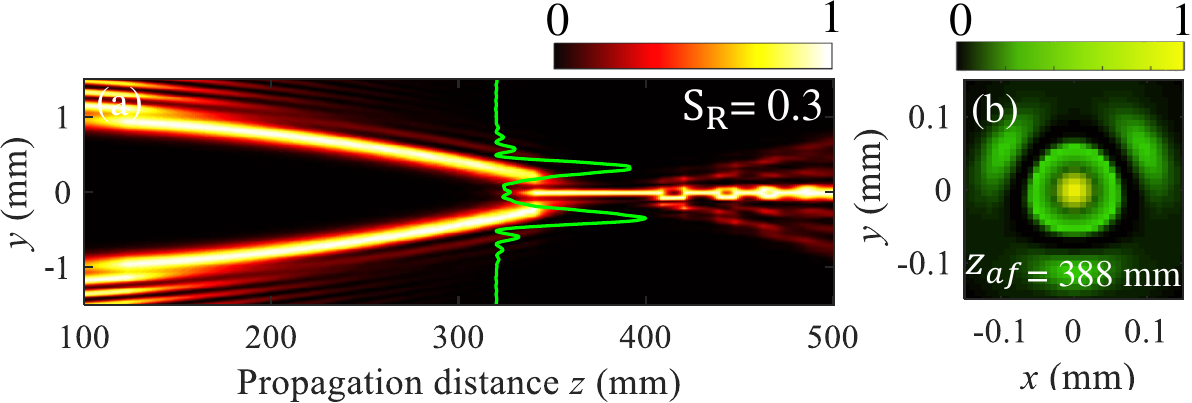}
\caption{Effect of vertical Trefoil aberration. In a turbulent media with $\mathrm{S_R}=0.3$ (a) cross-sectional intensity distribution of CADB as a function of propagation distance, and (b) the intensity distribution of CADB at $z_{af}=388$ mm. The solid green curve represents the intensity cross-section at $z=320$ mm. Note, Note, here turbulent media corresponds to a phase screen containing only vertical Trefoil aberration with certain amplitude.}
\label{fig17}
\end{figure}

Corresponding to turbulent media with $\mathrm{S_R}=0.7$, $\mathrm{S_R}=0.5$, and $\mathrm{S_R}=0.3$, the amplitudes ($a_{3,-3}$) of $Z_{3,-3}$ are found to be $0.05 ~\mu$m, $0.05 ~\mu$m, and $0.29 ~\mu$m, respectively (Fig.\,\ref{fig11}). As $a_{3,-3}$ is found to be appreciable only for the case of $\mathrm{S_R}=0.3$, so we have presented the results only for this case, as shown in Fig.\,\ref{fig17}. Figure\,\ref{fig17}(a) shows the cross-sectional intensity distribution of CADB as a function of propagation distance for turbulent media with $\mathrm{S_R}=0.3$. Figure\,\ref{fig17}(b) shows the intensity distribution of CADB at $z_{af}=388$ mm in the same turbulent media. As evident, at $z_{af}$ instead of getting a full radially symmetric focused beam (Fig.\,\ref{fig15}(a) and \ref{fig15}(c) for $\mathrm{S_{R}}=1$), the beam intensity becomes trefoil-shaped pattern (Fig.\,\ref{fig17}(b)). 

Furthermore, the oblique Trefoil aberration also introduces a threefold symmetric pattern into the wavefront, but it doesn't align with either the vertical or horizontal direction. Instead, the symmetry occurs in a direction tilted relative to these axes. 
\begin{figure}[htbp]
\centering
\includegraphics[height = 4.4cm, keepaspectratio = true]{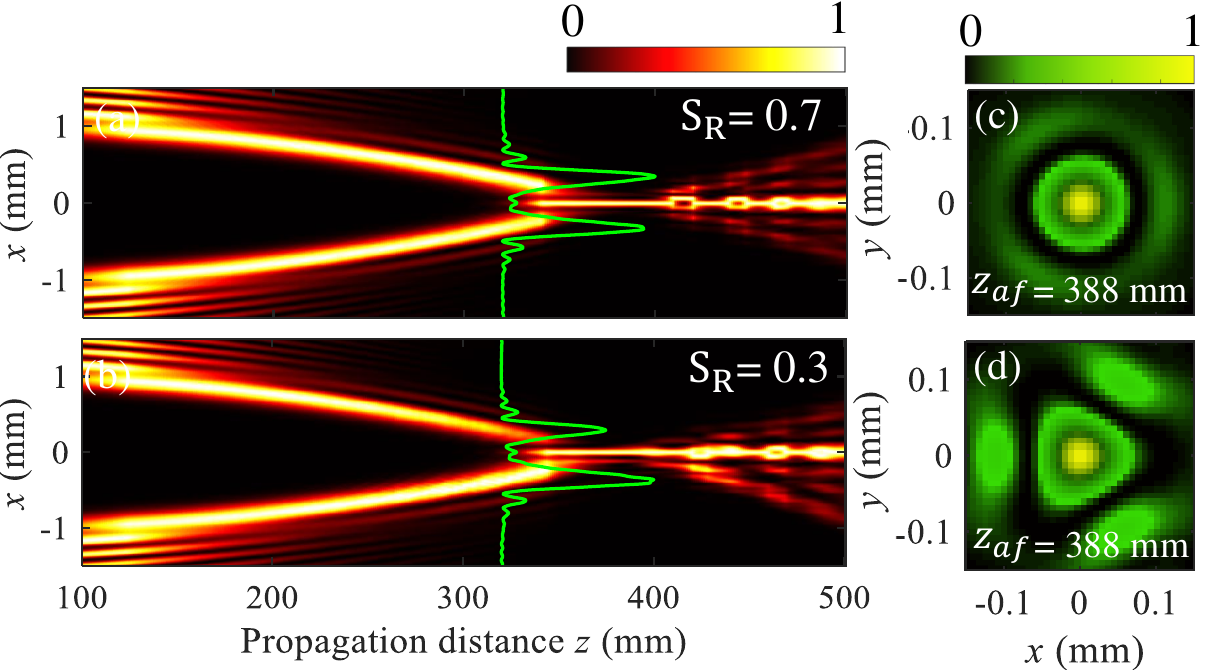}
\caption{Effect of oblique Trefoil aberration. The cross-sectional intensity distribution of CADB as a function of propagation distance in turbulent media with (a) $\mathrm{S_R}=0.5$, and (b) $\mathrm{S_R}=0.3$. The solid green curve represents the intensity cross-section at $z=320$ mm. Note, here turbulent media corresponds to a phase screen containing only oblique Trefoil aberration with certain amplitude. At $z_{af}=388$ mm, the intensity distribution of CADB for turbulent media with (c) $\mathrm{S_R}=0.5$, and (d) $\mathrm{S_R}=0.3$.}
\label{fig18}
\end{figure}
It is represented by $10^{\mathrm{th}}$ order Zernike polynomial, which is given as \cite{niu2022}
 \begin{equation}
 Z_{3,3}(r,\theta) = \sqrt{8}(r^3\cos3\theta).
 \end{equation} 
Corresponding to turbulent media with $\mathrm{S_R}=0.7$, $\mathrm{S_R}=0.5$, and $\mathrm{S_R}=0.3$, the amplitudes ($a_{3,3}$) of $Z_{3,3}$ are found to be $0.2 ~\mu$m, $-0.7 ~\mu$m, and $-1.23 ~\mu$m, respectively (Fig.\,\ref{fig11}). The results corresponding to $a_{3,3}=0.2 ~\mu$m ($\mathrm{S_R}=0.7$) and $a_{3,3}=-1.23 ~\mu$m ($\mathrm{S_R}=0.3$) are shown in Fig.\,\ref{fig18}.
As evident, the intensity of focused beam at $z_{af}$ turns into a trefoil-shaped pattern (three fold symmetric), which is tilted obliquely as compared to the case of vertical Trefoil aberration (Fig.\,\ref{fig17}(b)). The effect becomes more pronounced when $a_{3,3}$ increases to higher values. Note, the negative value of $a_{3,3}$ flips the intensity towards the opposite side (Fig.\,\ref{fig18}(a) and \ref{fig18}(b))(shown by the solid green curves).
\subsection{Primary Spherical aberration}
This aberration occurs when light rays, passing through the periphery of a lens or mirror, focus at different points than those passing through the center, which results in defocusing of images.
\begin{figure}[htbp]
\centering
\includegraphics[height = 3.60cm, keepaspectratio = true]{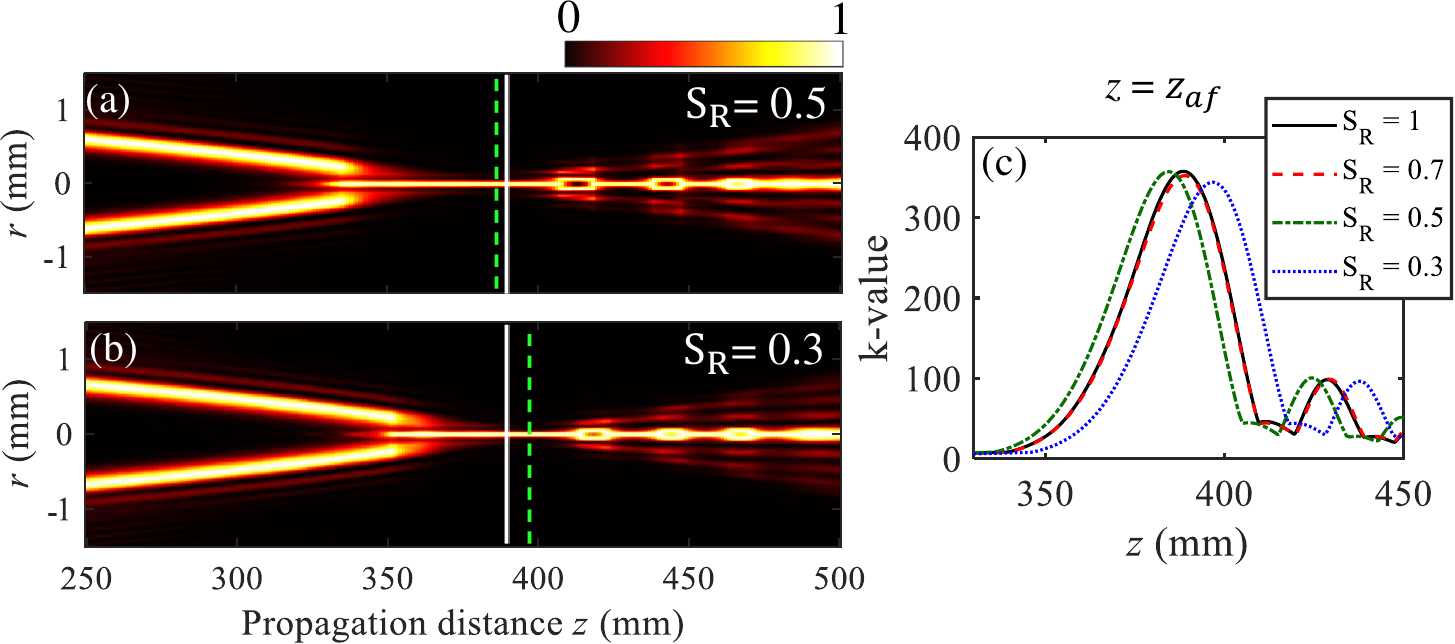}
\caption{Effect of primary Spherical aberration. The cross-sectional intensity distribution of CADB as a function of propagation distance for turbulent media with (a) $\mathrm{S_R}=0.5$, and (b) $\mathrm{S_R}=0.3$. The solid white and dashed green lines mark the locations of autofocusing position $z_{af}$ in free space ($\mathrm{S_R}=1$) and in turbulent media ($\mathrm{S_R}=0.5$ and $\mathrm{S_R}=0.3$), respectively. Note, here turbulent media corresponds to a phase screen containing only primary spherical aberration with certain amplitude. (c) The k-value as a function of propagation distance for $\mathrm{S_R}=1$ (solid black curve), $\mathrm{S_R}=0.7$ (dashed red curve), $\mathrm{S_R}=0.5$ (dot-dashed green curve) and $\mathrm{S_R}=0.3$ (dotted blue curve).}
\label{fig19}
\end{figure}
The primary spherical aberration is represented by $11^{\mathrm{th}}$ order Zernike polynomial, which is given as \cite{niu2022}
 \begin{equation}
 Z_{4,0}(r,\theta) = \sqrt{5}(6r^4-6r^2+1).
 \end{equation}
Corresponding to our considered turbulent media with $\mathrm{S_R}=0.7$, $\mathrm{S_R}=0.5$ and $\mathrm{S_R}=0.3$, the amplitudes ($a_{4,0}$) of this aberration are found to be $-0.141 ~\mu$m, $0.938 ~\mu$m, and $-1.12 ~\mu$m, respectively (Fig.\,\ref{fig11}). The results for primary spherical aberration induced effects on CADB are shown in Fig.\,\ref{fig19}. Figures\,\ref{fig19}(a) and \ref{fig19}(b) show the cross-sectional intensity distribution of CADB as a function of propagation distance for turbulent media with $\mathrm{S_R}=0.5$ and $\mathrm{S_R}=0.3$, respectively. Note, for $\mathrm{S_R}=0.7$ the amplitude $a_{4,0}$ is found to be relatively small, so the results are not included. Figure\,\ref{fig19}(c) shows the k-value as a function of propagation distance for $\mathrm{S_R}=1$ (solid black curve), $\mathrm{S_R}=0.7$ (dashed red curve), $\mathrm{S_R}=0.5$ (dot-dashed green curve), and $\mathrm{S_R}=0.7$ (dotted blue curve). As evident, for turbulent media with different $\mathrm{S_R}$ values, the autofocusing distance of CADB shifts by different amounts. In particular, the positive [negative] value of $a_{4,0}$ decreases [increases] $z_{af}$. Specifically, corresponding to values of $a_{4,0}=-0.141 ~\mu$m (for $\mathrm{S_R}=0.7$), $0.938 ~\mu$m (for $\mathrm{S_R}=0.5$), and $-1.12 ~\mu$m (for $\mathrm{S_R}=0.3$), the autofocusing distance $z_{af}$ is found to be 389 mm, 386 mm, and 397 mm, respectively (Figs.\,\ref{fig19}(a) and \ref{fig19}(b)).

 As shown in Fig.\,\ref{fig11}, the amplitudes of $12^{\mathrm{th}}$ to $29^{\mathrm{th}}$ Zernike coefficients are found to be relatively small, so we have not presented the results corresponding to these aberrations.  
\subsection{Tertiary x-Coma aberration}
In the considered turbulent media with different strengths, we have found an appreciable amount of tertiary x-Coma aberration, which is a higher order Coma aberration whose contribution provides additional complexity in aberration pattern that affects the image in a horizontal (x-axis) direction. It is is represented by $30^{\mathrm{th}}$ order Zernike polynomial, and is given as \cite{niu2022}
 \begin{equation}
 Z_{7,-1}(r,\theta) = 4(35r^7-60r^5+30r^3-4r)\cos\theta.
 \end{equation} 
\begin{figure}[htbp]
\centering
\includegraphics[height = 4.30cm, keepaspectratio = true]{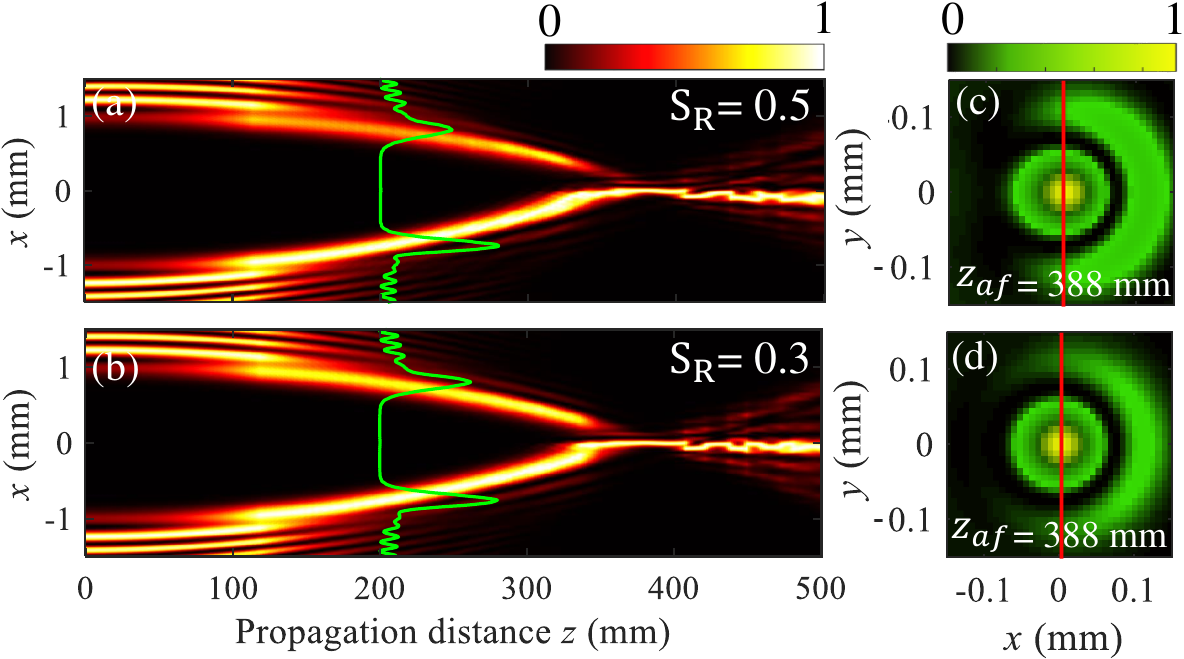}
\caption{Effect of tertiary x-Coma aberration. The cross-sectional intensity distribution of CADB as a function of propagation distance in turbulent media with (a) $\mathrm{S_R}=0.5$, and (b) $\mathrm{S_R}=0.3$. The solid green curve represents the intensity cross-section at $z=200$ mm. Note, here turbulent media corresponds to a phase screen containing only tertiary x-Coma aberration with certain amplitude. At $z_{af}=388$ mm, the intensity distribution of CADB in turbulent media with (c) $\mathrm{S_R}=0.5$, and (d) $\mathrm{S_R}=0.3$.}
\label{fig20}
\end{figure}

In considered turbulent media with $\mathrm{S_R}=0.7$, $\mathrm{S_R}=0.5$, and $\mathrm{S_R}=0.3$, the amplitudes ($a_{7,-1}$) of $Z_{7,-1}$ are found to be $-0.01 ~\mu$m, $0.342 ~\mu$m, and $0.176 ~\mu$m, respectively. For $\mathrm{S_R}=0.5$ ($a_{7,-1}=0.342 ~\mu$m) and $\mathrm{S_R}=0.3$ ($a_{7,-1}=0.176 ~\mu$m), the results are shown in Fig.\,\ref{fig20}. Figures\,\ref{fig20}(a) and \ref{fig20}(b) show the cross-sectional intensity distribution of CADB as a function of propagation distance for turbulent media with $\mathrm{S_R}=0.5$, and $\mathrm{S_R}=0.3$, respectively. The solid green curve represents the intensity cross-section at $z=200$ mm. Figures\,\ref{fig20}(c) and \ref{fig20}(d) show the intensity distribution of CADB at $z_{af}=388$ mm for $\mathrm{S_R}=0.5$, and $\mathrm{S_R}=0.3$, respectively. The red line marks the reference position at $x=0$. As evident, unlike Coma aberration (Sec.\,\ref{comaabr}) there is no apparent beam wandering of focused beam spot at $z_{af}=388$ mm (marked by red line). However, this aberration leads to asymmetric intensity distribution along x-axis, similar to horizontal Coma aberration (Fig.\,\ref{fig16}).

Further, the amplitudes of Zernike coefficients with orders $>30$ are found to be relatively small (Fig.\,\ref{fig11}), so we have not presented the results corresponding to those aberrations.  

The Zernike polynomial based analysis clearly suggests that when CADB propagates through the turbulent media, the adverse effects are caused by the presence of different types of aberrations. Particularly, the beam wandering occurs as a result of y-tilt, vertical Coma, horizontal Coma, and tertiary x-Coma aberrations. A shift in the autofocusing distance occurs due to the presence of defocus and primary spherical aberrations. The splitting of rings into lobe-like structures is caused by vertical trefoil and oblique trefoil aberrations, while the directional shift in intensity is predominantly due to Coma and Trefoil aberrations. Overall, the combined effect of these aberrations influences the autofocusing abilities of CADB to certain extent, as shown above in Sec.\,\ref{results}. This analysis can be helpful in improving further the propagation and autofocusing properties of CADB through the turbulent media. 
\section{Conclusions}\label{concl}
We have experimentally and numerically investigated the abruptly autofocusing properties of CADBs in free space ($\mathrm{S_R}=1$) and in turbulent media with varying strength, namely, low turbulence ($\mathrm{S_R}=0.7$), medium turbulence ($\mathrm{S_R}=0.5$), and high turbulence ($\mathrm{S_R}=0.3$). The investigations are performed for a single realization as well as multiple realizations of turbulent media. The results are quantified by the k-value, FWHM of k-value curve, autofocusing distance $z_{af}$, focal spot size and diffraction efficiency. We have found that abrupt autofocusing of CADB remains present even when the CADB is propagated in a relatively strong turbulent media. The maximum k-value decreases with increasing the strength of turbulence. The FWHM of k-value curves does not change significantly with increasing the strength of turbulence, indicating that abruptness in the autfocusing of CADB remains reasonably well. We have found that autofocusing distance changes by small amount with increasing the strength of turbulence. At $z_{af}$, the focal spot size of CADB grows gradually with increasing the strength of turbulent media, and particularly for $\mathrm{S_{R}}=0.3$ (strong turbulent media) it grows by a factor of 2.2, indicating reasonably good focusing behaviour. The diffraction efficiency of CADB changes only by a factor of 1.7 when $\mathrm{S_{R}}$ is varied from 1 (no turbulence) to 0.3 (strong turbulence), indicating good confinement of intensity in strong turbulent media. Further, the spatial distortions in CADB caused by turbulence are quantified by an overlap integral, which shows that CADB possesses reasonably good resilience against the turbulence.

Further, we have also performed a detailed comparison between the the CADB and conventional Gaussian beam with and without plano-convex lens. We have found that CADBs possess superior properties (abrupt autofocusing, maximum k-value, focal spot size and diffraction efficiency) than Gaussian beams in turbulent media. To understand the adverse effects caused by turbulent media on the propagation of CADB, we have carried out detailed investigations based on Zernike polynomials. This analysis reveals that different kinds of aberrations present in turbulent media leads to distortions in the spatial structure as well as other properties of CADBs. To compensate for the effects of turbulence, it is essential to identify the specific aberrations present and their magnitudes. This investigation allows us to utilize these beams more effectively in applications.

The ability of CADBs to show abrupt autofocusing, even in highly turbulent media, demonstrates their great robustness. As a result, these beams prove to be highly valuable for a variety of applications, particularly in fields such as biomedical treatment, where CADBs need to travel through disordered media, high-resolution imaging, efficient guidance, and trapping of microparticles, where their resilience against turbulence becomes highly required.

\begin{acknowledgments}
We acknowledge the funding support from Indian Institute of Technology Ropar and Science and Engineering Research Board (Grant no. CRG/2021/003060). Anita Kumari acknowledges the fellowship support from University Grants Commission (UGC).

\end{acknowledgments}
\appendix
\section{Effect of multiple realizations of turbulent media} \label{appenA}
In our investigations, we have realized turbulent media of a desire length and strength by a phase screen constructed from a complex random matrix ($R_{NN}$). The complex random matrix can have several configurations while keeping the strength of randomness fixed. Further, the turbulent media can be dynamic in nature, where randomness can change locally, however, the overall turbulence strength remains the same. Therefore, for the generalizations of our study, we have performed statistical analysis of maximum k-value, FWHM of k-value curve and autofocusing distance $z_{af}$, over 80 realizations of turbulent phase screens. Note, the results remains almost the same for realizations $>80$. 
\begin{figure}[htbp]
\centering
\includegraphics[height = 4.50cm, keepaspectratio = true]{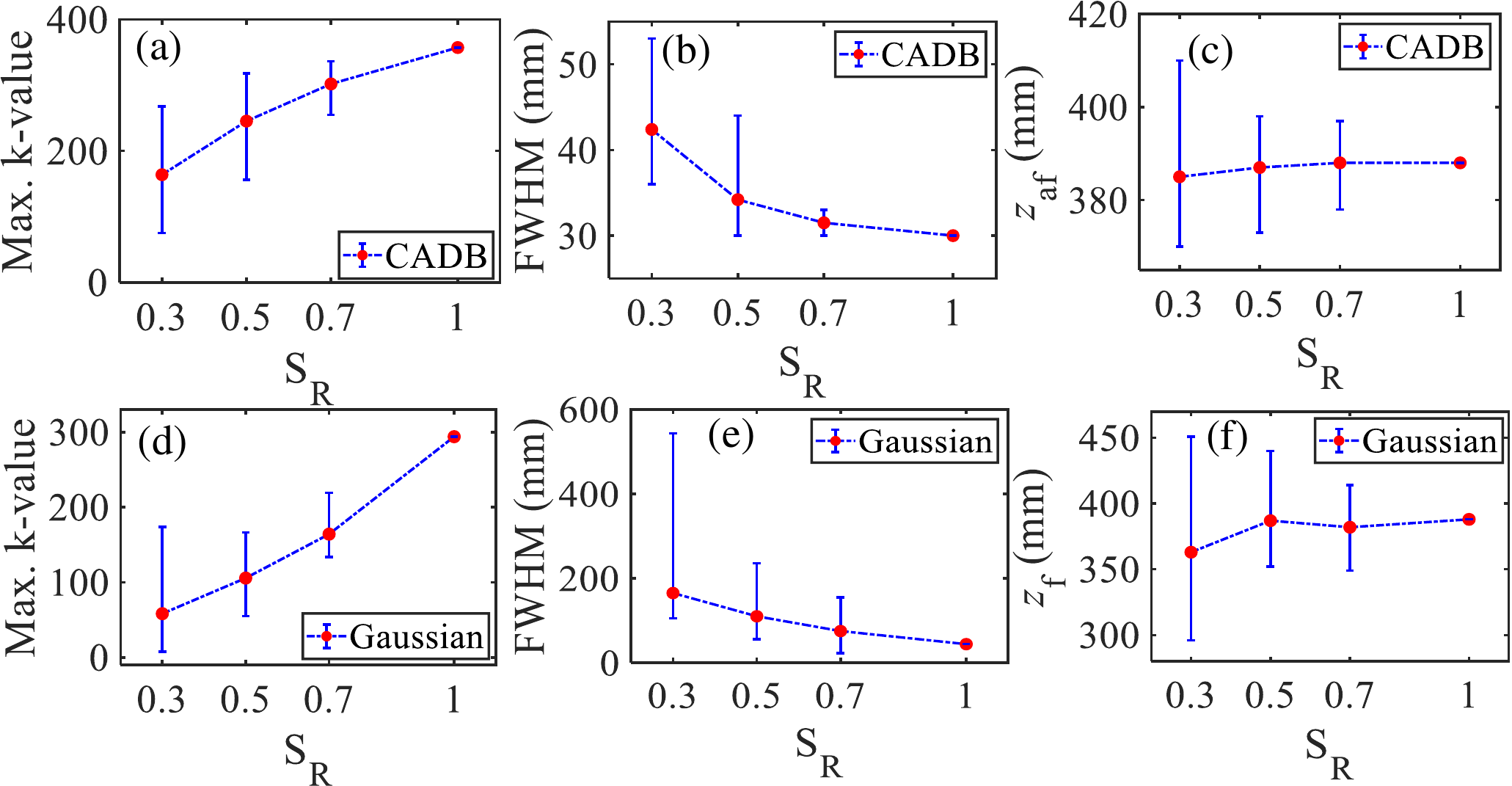}
\caption{Simulated results over 80 realizations of turbulent media (turbulent phase screens). The comparison between the CADB (upper row) and Gaussian beam with a plano-convex lens (lower row) at different turbulent strengths. With respect to $\mathrm{S_R}$, the variation of (a,d) maximum k-value, (b,e) FWHM of k-value curve, and (c,f) autofocusing distance $z_{af}$. The error bar represents the variation of quantity over 80 realizations of turbulent media with a fixed $\mathrm{S_R}$. The filled red circles denote the average value of 80 realizations.}
\label{fig23}
\end{figure}

The results for CADB (top row) and Gaussian (bottom row) are presented in Fig.\,\ref{fig23}. Note, for a comparison with autofocused CADB, we have propagated a Gaussian beam through a plano-convex lens with focal length equivalent to autofocusing distance of CADB. Figures\,\ref{fig23}(a)-\ref{fig23}(c) and Figs.\,\ref{fig23}(d)-\ref{fig23}(f) show the variation of maximum k-value, FWHM of k-value curve and autofocusing distance $z_{af}$ at different strength of turbulent media, for CADB and Gaussian beam, respectively. The error bar denotes the range within which the values are varying for different realizations of turbulent media with a fixed $\mathrm{S_R}$ value. The filled red circles denote the average value of 80 realizations. It is evident that over different realizations of turbulent media, the values vary in a certain range, and particularly with increasing the strength of turbulence, the range of variation gets extended. More specifically, for 80 realizations of turbulent media with $\mathrm{S_R}=0.7$, we have found the variation of $z_{af}$ in between [378 mm$-$397 mm], maximum k-value in between [$350-255$], and FWHM of k-value curve in between [30 mm$-$33 mm]. Similarly, for $\mathrm{S_R}=0.5$, $z_{af}$ varies in between [373 mm$-$398 mm], maximum k-value varies in between [$318-156$], and FWHM of k-value curve varies in between [30 mm$-$44 mm]. For $\mathrm{S_R}=0.3$, the $z_{af}$ varies in between [370 mm$-$410 mm], maximum k-value varies in between [$268-75$], and FWHM of k-value curve varies in between [36 mm$-$53 mm].

\section{Propagation of Gaussian beam in turbulent media}\label{appenB}
Here we have investigated the propagation of a Gaussian beam in the same turbulent media for which the propagation of CADBs is analyzed. As CADB possesses the autofocusing features, so for a comparison between the Gaussian beam and CADB, we have propagated Gaussian beam with and without a focusing lens. With a focusing lens, the aim is to compare the focusing abilities in turbulent media of different strengths.  
\begin{figure*}[htbp]
\centering
\includegraphics[height = 8.70cm, keepaspectratio = true]{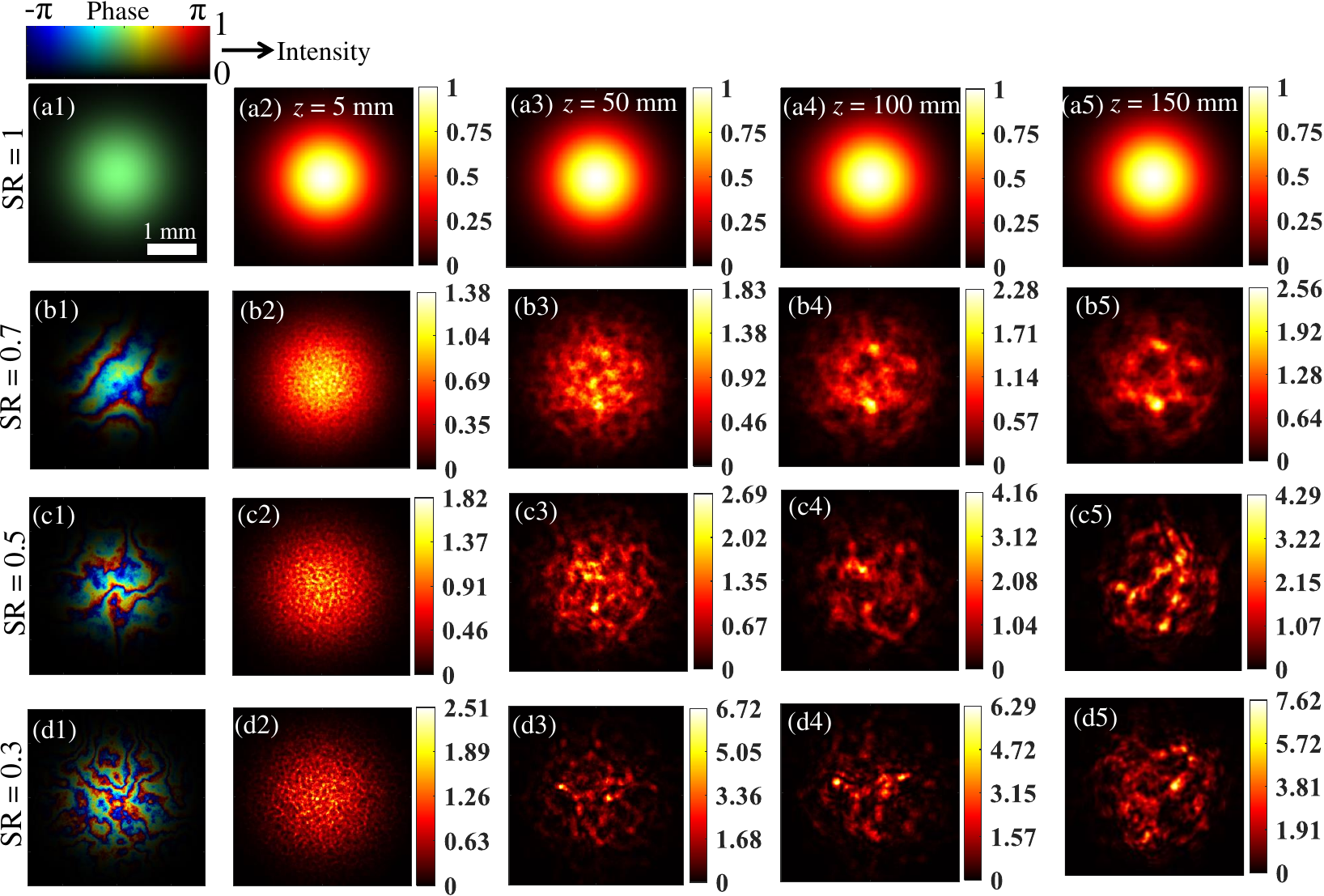}
\caption{Simulated results. The propagation of a Gaussian beam in (a1-a5) free space ($\mathrm{S_{R}}=1$), (b1-b5) turbulent media with $\mathrm{S_{R}}=0.7$, (c1-c5) turbulent media with $\mathrm{S_{R}}=0.5$, and (d1-d5) turbulent media with $\mathrm{S_{R}}=0.3$. Note, the first column shows complex valued field of Gaussian beams (a1, b1, c1, and d1), where ``color" represents the ``phase" and ``contrast" as ``intensity".  The beam waist of Gaussian beam is taken as $1.5$ mm.}
\label{fig21}
\end{figure*}
\begin{figure*}[htbp]
\centering
\includegraphics[height = 8.70cm, keepaspectratio = true]{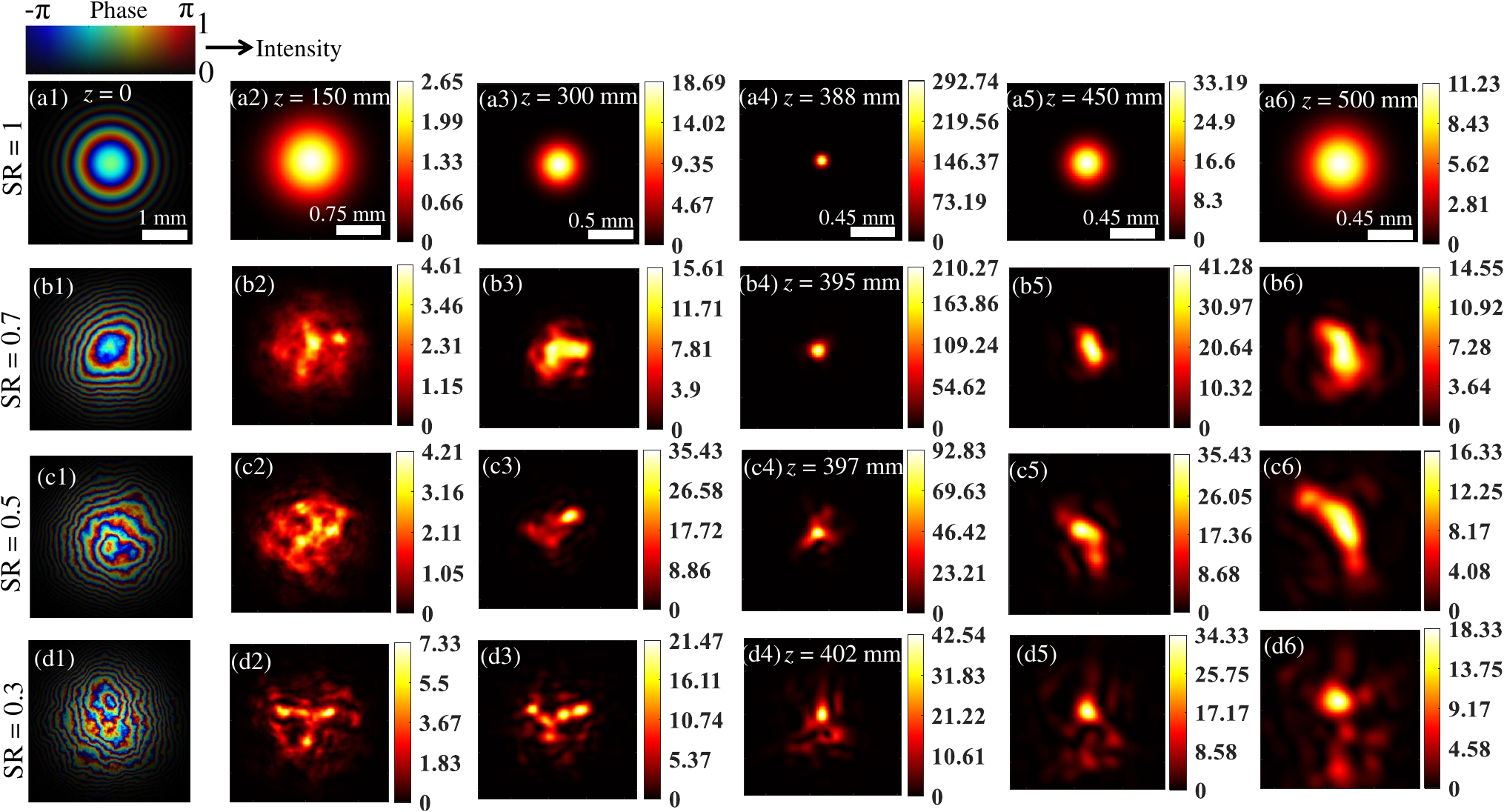}
\caption{The focusing of a Gaussian beam with a plano-convex lens (focal length $f=388$ mm) in (a1-a6) free space ($\mathrm{S_{R}}=1$), (b1-b6) turbulent media with $\mathrm{S_{R}}=0.7$, (c1-c6) turbulent media with $\mathrm{S_{R}}=0.5$, and (d1-d6) turbulent media with $\mathrm{S_{R}}=0.3$. Note, the first column shows complex valued field of Gaussian beams (a1, b1, c1, and d1), where ``color" represents the ``phase" and ``contrast" as ``intensity".  The beam waist of Gaussian beam is taken as $1.5$ mm.}
\label{fig22}
\end{figure*}

We consider a Gaussian beam with a $1.5$ mm beam waist (comparable to CADB's size) and propagate it with and without focusing lens in free space ($\mathrm{S_R}=1$) and turbulent media with $\mathrm{S_R}=0.7,~0.5,~\mathrm{and}~0.3$. The results on the propagation of a Gaussian beam without focusing lens are shown in Fig.\,\ref{fig21}. The intensity distribution at different propagation distances is shown in free space with $\mathrm{S_R}=1$ (Figs.\,\ref{fig21}(a1)-\ref{fig21}(a5)), and turbulent media with $\mathrm{S_R}=0.7$ (Figs.\,\ref{fig21}(b1)-\ref{fig21}(b5)), $\mathrm{S_R}=0.5$ (Figs.\,\ref{fig21}(c1)-\ref{fig21}(c5)), and $\mathrm{S_R}=0.3$ (Figs.\,\ref{fig21}(d1)-\ref{fig21}(d5)). Note, in Fig.\,\ref{fig21}, the first column represents the complex-valued field of the Gaussian beam, where ``color" represents the ``phase" and ``contrast" as ``intensity". As shown, Gaussian beam in free space remains a Gaussian, however, the size increases with increasing the distance due to diffraction. When Gaussian beam propagates in turbulent media with $\mathrm{SR=0.7}$, it undergoes strong distortions and converts into speckled intensity distribution (Figs.\,\ref{fig21}(b1)-\ref{fig21}(b5)). Upon increasing the strength of turbulence, the distortions become more strong, and the beam changes into more pronounced speckle intensity distribution (speckle size reduces) (Figs.\,\ref{fig21}(c1)-\ref{fig21}(c5) and Figs.\,\ref{fig21}(d1)-\ref{fig21}(d5)). This shows that the Gaussian beam is highly sensitive to turbulent media.

For a close comparison of the Gaussian beam with CADBs, we propagate the Gaussian beam through a plano-convex lens of focal length $f=388$ mm. This ensures that Gaussian beam focuses at a distance $z=388$ mm in free space (equivalent to autofocusing distance of CADB in free space). The results of focusing of Gaussian beam in free space ($\mathrm{S_R}=1$) and in turbulent media with $\mathrm{S_R}=0.7,~0.5,~ \mathrm{and}~ 0.3$ are shown in Fig.\,\ref{fig22}. Note, in Fig.\,\ref{fig22}, the first column represents the complex-valued field of the Gaussian beam with a plano-convex lens, where ``color" represents the ``phase" and ``contrast" as ``intensity". The intensity distribution of Gaussian beam at different propagation distances is shown in free space with $\mathrm{S_R}=1$ (Figs.\,\ref{fig22}(a1)-\ref{fig22}(a6)), and in turbulent media with $\mathrm{S_R}=0.7$ (Figs.\,\ref{fig22}(b1)-\ref{fig22}(b6)), $\mathrm{S_R}=0.5$ (Figs.\,\ref{fig22}(c1)-\ref{fig22}(c6)), and $\mathrm{S_R}=0.3$ (Figs.\,\ref{fig22}(d1)-\ref{fig22}(d6)). In free space, Gaussian beam focuses at a distance $z=388$ mm, where intensity becomes tightly focused in the form of a high intensity peak, and after that it diverges (Figs.\,\ref{fig22}(a1)-\ref{fig22}(a6)). In a turbulent media with low strength $\mathrm{S_R}=0.7$, Gaussian beam experiences significant distortions upon propagation, however, it still shows autofocusing at a shifted distance $z=395$ mm with increased spot size (Figs.\,\ref{fig22}(b1)-\ref{fig22}(b6)). Upon increasing the strength of turbulent media to $\mathrm{S_R}=0.5$ and $\mathrm{S_R}=0.3$, Gaussian beam experiences more pronounced distortions upon propagation, and shows reduced focusing abilities (significant intensity lies in the background) at $z=397$ mm and 402 mm, respectively (Figs.\,\ref{fig22}(c1)-\ref{fig22}(c6) and Figs.\,\ref{fig22}(d1)-\ref{fig22}(d6)). Further, the k-value (maximum value in intensity color bar) considerably decreases with increasing the strength of turbulent media.

%
\end{document}